\documentclass[aps,prd,groupedaddress,showpacs,showkeys]{revtex4}
\usepackage{amssymb}
\usepackage{amsmath}
\usepackage{amsfonts}
\usepackage{epsfig}
\usepackage{graphicx}
\usepackage{tabularx}

\newcommand{\beaa}{\begin{eqnarray*}}
\newcommand{\enaa}{\end{eqnarray*}}
\newcommand{\bea}{\begin{eqnarray}}
\newcommand{\ena}{\end{eqnarray}}
\newcommand{\seq}{\begin{subequations}}
\newcommand{\sen}{\end{subequations}}
\newcommand{\eq}{\begin{eqnarray}}
\newcommand{\en}{\end{eqnarray}}

\def\shiftdown#1{#1\llap{\lower.04ex\hbox{#1}}}

\newcommand{\ra}{\rangle}
\newcommand{\la}{\langle}

\def\arraystretch{1.5}

\begin{document}

\title{Nucleon resonances  in AdS/QCD} 

\author{
Thomas Gutsche$^1$,
Valery E. Lyubovitskij$^1$
\footnote{On leave of absence
from Department of Physics, Tomsk State University,
634050 Tomsk, Russia},
Ivan Schmidt$^2$,
Alfredo Vega$^3$
\vspace*{1.2\baselineskip}\\
}

\affiliation{
$^1$ Institut f\"ur Theoretische Physik,
Universit\"at T\"ubingen, \\
Kepler Center for Astro and Particle Physics,
\\ Auf der Morgenstelle 14, D-72076 T\"ubingen, Germany
\vspace*{1.2\baselineskip} \\
\hspace*{-1cm}
$^2$ Departamento de F\'\i sica y Centro Cient\'\i fico 
Tecnol\'ogico de Valpara\'\i so (CCTVal), Universidad T\'ecnica
Federico Santa Mar\'\i a, Casilla 110-V, Valpara\'\i so, Chile
\vspace*{1.2\baselineskip} \\
$^3$ Departamento de F\'isica y Astronom\'ia, \\
Universidad de Valpara\'iso,\\
A. Gran Breta\~na 1111, Valpara\'iso, Chile
\vspace*{1.2\baselineskip} \\
}

\date{\today}

\begin{abstract}

We describe the electroproduction of the $N(1440)$ Roper resonance 
in soft-wall AdS/QCD. The Roper resonance is identified as
the first radially excited state of the nucleon, where higher Fock states
in addition to the three-quark $(3q)$ component are included. 
The main conclusion is that the leading $3q$ component plays the dominant 
role in the description of electroproduction properties of this resonance:
form factors, helicity amplitudes, and charge densities.  
The obtained results are in good agreement with the recent results of 
the CLAS Collaboration at JLab. 

\end{abstract}



\maketitle

\section{Introduction}

The study of electromagnetic properties of nucleon resonances, such as  
the Roper resonance, opens new opportunities for understanding the
structure of hadrons. One of the more promising experiments of this type 
has been performed by the CLAS Collaboration 
at JLab~\cite{Aznauryan:2009mx,:2012sha}, which will be continued at 
the upgraded JLab facilities with a 12-GeV energy beam~\cite{Dudek:2012vr}. 
This calls for a comprehensive theoretical analysis of this process. 
From the theoretical side, starting from 1980, different types of 
approaches/models [nonrelativistic and relativistic quark models, 
potential and hadronic molecular approaches, Dyson-Schwinger equation 
framework, light-front holographic quantum chromodynamics (QCD), etc.] 
have been proposed and developed for the description of electroexcitations 
of nucleon resonances~\cite{theory}-\cite{Aznauryan:2012ba} (for recent 
reviews see  e.g.~\cite{Obukhovsky:2011sc,Aznauryan:2011qj,Aznauryan:2012ba}). 
For example, in some recent theoretical 
developments~\cite{Obukhovsky:2011sc,Aznauryan:2012ec} 
it is pointed out that a realistic description of the current data on Roper 
electroproduction needs to include additional degrees of freedom for 
this state, such as a nucleon-scalar $\sigma$ meson molecular component. 
In Ref.~\cite{deTeramond:2011qp} the Dirac form factor for the 
electromagnetic nucleon-Roper transition has been calculated in 
light-front holographic QCD. 

In this paper we consider Roper electroproduction in a soft-wall 
AdS/QCD model~\cite{Approach,Approach2,Gutsche:2012bp}, which not only 
includes the leading three-quark ($3q$) state but also higher Fock components. 
In Ref.~\cite{Gutsche:2012bp} we proposed  this AdS/QCD model 
as an approach to baryon structure, and successfully applied it to 
the study of nucleon electromagnetic form factors in the Euclidean region 
of transverse momentum squared up to 30 GeV$^2$. 
In~\cite{Gutsche:2012bp} we found that the inclusion of higher Fock states 
is relevant for the quantitative reproduction of baryons properties 
--- masses, and the electromagnetic form factors at small $Q^2$ including 
the electromagnetic radii.  
We truncated the tower of Fock states to the twist 
dimension 
$\tau=5$, because the contribution of higher Fock states 
to the hadronic form factors scales as $(1/Q^2)^{\tau-1}$ at higher $Q^2$ and 
is therefore suppressed. Another reason for the truncation to $\tau=5$ was 
to reduce the number of free parameters. 

The paper is structured as follows. First, in
Sec.~II, we briefly discuss the basic notions of the approach. In
Secs.~III and IV we consider applications of our approach 
to the electroproduction of the Roper resonance. 
Finally, in Sec.~V, we summarize our results.

\section{Approach}

\subsection{Action for the spin $J=\frac{1}{2}$ fermion field in AdS space}

Here we briefly review our approach. First, we specify the
five-dimensional AdS metric: 
\eq
ds^2 = 
g_{MN} dx^M dx^N &=& \eta_{ab} \, e^{2A(z)} \, dx^a dx^b = e^{2A(z)}
\, (\eta_{\mu\nu} dx^\mu dx^\nu - dz^2)\,, 
\nonumber\\
\eta_{\mu\nu} &=& {\rm diag}(1, -1, -1, -1, -1) \,,
\en
where $M$ and $N = 0, 1, \cdots , 4$ are the space-time (base manifold) 
indices, $a=(\mu,z)$ and $b=(\nu,z)$ are the local Lorentz (tangent) indices,
and $g_{MN}$ and  $\eta_{ab}$ are curved and flat metric tensors,  
respectively, which
are related by the vielbein $\epsilon_M^a(z)= e^{A(z)} \, \delta_M^a$ as 
$g_{MN} =\epsilon_M^a \epsilon_N^b \eta_{ab}$. Here
$z$ is the holographic coordinate, $R$ is the AdS radius, and $g =
|{\rm det} g_{MN}| = e^{10 A(z)}$. In the following we restrict
ourselves to a conformal-invariant metric with $A(z) = \log(R/z)$.

The relevant AdS/QCD action for the fermion field of twist $\tau$ 
is~\cite{Gutsche:2012bp} 
\eq\label{eff_action} 
S_\tau = \int d^4x dz \, \sqrt{g} \, e^{-\varphi(z)} \, \sum\limits_{i=+,-} 
\bar\Psi_{i,\tau}(x,z) 
\, \hat{\cal D}_i(z) \, \Psi_{i,\tau}(x,z) \,, 
\en 
where 
\eq 
\hat{\cal D}_\pm(z) &=&  \frac{i}{2} \Gamma^M 
\! \stackrel{\leftrightarrow}{\partial}_{_M} 
\, \mp \,  (\mu + U_F(z))\,,
\en 
and $\Psi_{\pm,\tau}(x,z)$ is the pair of bulk fermion fields, which 
are the holographic analogues of the left- and right-chirality operators in 
the 4D theory.  
Here $A \! \stackrel{\leftrightarrow}{\partial} \! B
\equiv A (\partial B) - (\partial A) B$, 
$\varphi(z) = \kappa^2 z^2$ is the dilaton field with $\kappa$ being 
a free scale parameter. $\Gamma^M = \epsilon_a^M \Gamma^a$ and  
$\Gamma^a=(\gamma^\mu, - i\gamma^5)$ are the five-dimensional Dirac matrices 
(we use the chiral representation for the $\gamma^\mu$ and $\gamma^5$ 
matrices; see details in Refs.~\cite{Approach,Gutsche:2012bp}).  
The quantity $\mu$ is the bulk fermion mass related to the scaling dimension 
$\tau$ as $m = \mu R = \tau - 3/2$. Note that the scaling
dimension of the AdS fermion field is holographically identified 
with the scaling dimension of the baryon interpolating
operator $\tau = N + L$; 
$N$ is the number of partons in the baryon and 
$L = {\rm max} \, | L_z |$ is the maximal 
value of the $z$ component of the quark orbital angular momentum 
in the light-front wave function~\cite{Soft_wall2aa,Soft_wall2b}.  
$U_F(z) = \varphi(z)/R$ is the effective potential depending  
on the dilaton field. Its presence is necessary for the following reason.  
The form of the potential $U_F(z)$ is constrained in order to
get solutions of the equations of motion (EOMs) for the fermionic 
Kaluza-Klein (KK) modes of left and right chirality, and to have the 
correct asymptotic behavior of the nucleon electromagnetic form factors 
at large $Q^2$~\cite{Approach,Gutsche:2012bp,Soft_wall6,Soft_wall7}. 

Notice that the fermion masses $m$ and the effective potentials $U_F(z)$ 
corresponding to the fields $\Psi_+$ and $\Psi_-$ have opposite 
signs according to the $P$-parity transformation 
(see details in Ref.~\cite{Gutsche:2012bp}). 
The absolute sign of the fermion mass is related to the 
chirality of the boundary operator~\cite{Baryons_ADS_QCD1,Baryons_ADS_QCD2}.  
According to our conventions the QCD operators ${\cal O}_R$ and ${\cal O}_L$ 
have positive and negative chirality, and therefore the mass 
terms of the bulk fields $\Psi_+$ and $\Psi_-$ have absolute 
signs ``plus'' and ``minus'', respectively. 
The fields $\Psi_\tau$ describe the AdS fermion 
field with different scaling dimension: $\tau = 3, 4, 5$, etc. 

In Ref.~\cite{Gutsche:2012bp} we demonstrated that our soft-wall holographic 
model reproduces the main features of the electromagnetic structure
of the nucleon. In particular, we gained the following results:
the analytical power scaling of the elastic nucleon
form factors at large momentum transfers in accordance with quark-counting
rules; reproduction of experimental data for magnetic moments and
electromagnetic radii.

\subsection{Mass spectrum} 

One advantage of the soft-wall AdS/QCD model is that most of the 
calculations can be done analytically. 
Here we show how in this approach the baryon spectrum and 
wave functions are generated following the procedure presented in
Refs.~\cite{Soft_wall6,Soft_wall7,Approach,Gutsche:2012bp}.  
First, we rescale the fermionic fields as
\eq
\Psi_{i, \tau}(x,z) = e^{\varphi(z)/2} \psi_{i, \tau}(x,z)  
\en
and remove the dilaton field from the overall exponential.
In terms of the field $\psi_\tau(x,z)$ the modified action in 
the Lorentzian signature reads as 
\eq\label{S_F2}
S_\tau = \int d^4x dz \, e^{4A(z)} \, \sum\limits_{i=+,-} 
\, \bar\psi_{i,\tau}(x,z) \, 
\biggl\{ i\not\!\partial + \gamma^5\partial_z
+ 2 A^\prime(z) \gamma^5
-  \delta_i \frac{e^{A(z)}}{R} \Big(m + \varphi(z)\Big)  \biggr\} 
\, \psi_{i, \tau}(x,z)\,,
\en
where $\not\!\partial = \gamma^\mu \, \partial_\mu$, $\delta_\pm = \pm 1$.
The fermion field $\psi_{i, \tau}(x,z)$ satisfies the following 
EOM~\cite{Soft_wall6,Soft_wall7,Approach,Gutsche:2012bp}: 
\eq
\biggl[ i\not\!\partial + \gamma^5\partial_z
+ 2 A^\prime(z) \gamma^5
\mp \frac{e^{A(z)}}{R} \Big(m + \varphi(z)\Big) \biggr] 
\psi_{\pm,\tau}(x,z) = 0\,. 
\en

Next we split the fermion field into left- and right-chirality
components 
\eq
\psi_{i, \tau}(x,z) = \psi^L_{i,\tau}(x,z) + \psi^R_{i, \tau}(x,z)\,, \quad
\psi^{L/R}_{i, \tau}(x,z) = 
\frac{1 \mp \gamma^5}{2} \psi_{i, \tau}(x,z) \,, \quad
\gamma^5 \psi^{L/R}_{i, \tau}(x,z) = \mp \psi^{L/R}_{i, \tau}(x,z) \,,
\en
and perform a KK expansion for the $\psi^{L/R}_{i, \tau}(x,z)$ fields 
\eq
\psi^{L/R}_{i, \tau}(x,z) = \frac{1}{\sqrt{2}} \, \sum\limits_n
\ \psi^{L/R}_n(x) \ F^{L/R}_{i, \tau, n}(z) \,, 
\en
where $\psi^{L/R}_n(x)$ are the four-dimensional boundary 
fields (KK modes). These are Weyl spinors forming the 
Dirac bispinors $\psi_n(x) = \psi^L_n(x) + \psi^R_n(x)$, 
and $F^{L/R}_{i, \tau,n}(z)$ are the normalizable profile  
functions. Due to four-dimensional $P$- and $C$-parity invariance
the bulk profiles are related as~\cite{Gutsche:2012bp}:
\eq 
F^R_{\pm, \tau, n}(z) = \pm F^L_{\mp, \tau, n}(z)\,. 
\en 
Using this constraint, in the following we use the simplified notations: 
\eq 
F^R_{\tau, n}(z) &\equiv& 
F^R_{+, \tau, n}(z) = F^L_{-, \tau, n}(z)\,, \nonumber\\ 
F^L_{\tau, n}(z) &\equiv& 
F^L_{+, \tau, n}(z) = - F^R_{-, \tau, n}(z)\,. 
\en 
Note that the profiles $F^{L/R}_{\tau, n}(z)$ are the 
holographic analogues of the nucleon wave functions with 
specific radial quantum number $n$ and twist dimension $\tau$ 
(the latter corresponds to the specific partonic content of the nucleon Fock 
component), which  satisfy
the two coupled one-dimensional EOMs~\cite{Gutsche:2012bp}: 
\eq
\biggl[\partial_z \pm \frac{e^{A}}{R} \, \Big(m+\varphi\Big)
+ 2 A^\prime \biggr] F^{L/R}_{n, \tau}(z) = \pm M_{n\tau}
F^{R/L}_{n, \tau}(z) \,. 
\en
Therefore, the main aim is to find solutions for the 
bulk profiles of the AdS field in the $z$ direction and then calculate 
the physical properties of hadrons. 
After straightforward algebra one can obtain the decoupled EOMs: 
\eq
\biggl[ -\partial_z^2 - 4 A^\prime \partial_z
+ \frac{e^{2A}}{R^2} (m+\varphi)^2
\mp \frac{e^{A}}{R} \Big(A^\prime (m+\varphi) + \varphi^\prime\Big)
- 4 A^{\prime 2} - 2 A^{\prime\prime}
\biggr] F^{L/R}_{\tau, n}(z) = M_{n\tau}^2 F^{L/R}_{\tau, n}(z) \,.
\en
Performing the substitution
\eq
F^{L/R}_{\tau, n}(z) = e^{- 2 A(z)} \, f^{L/R}_{\tau, n}(z)
\en
we derive the Schr\"odinger-type EOM for $f^{L/R}_{\tau, n}(z)$
\eq\label{eq_KK} 
\biggl[ -\partial_z^2
+ \frac{e^{2A}}{R^2} (m+\varphi)^2
\mp \frac{e^{A}}{R} \Big(A^\prime (m+\varphi) +
 \varphi^\prime\Big) \biggr] f^{L/R}_{\tau, n}(z) 
= M_{n\tau}^2 \, f^{L/R}_{\tau, n}(z) \,.
\en
For $A(z)=\log(R/z)$, $\varphi(z)=\kappa^2 z^2$ we get
\eq
\biggl[ -\partial_z^2
+ \kappa^4 z^2 + 2 \kappa^2 \Big(m \mp \frac{1}{2} \Big)
+ \frac{m (m \pm 1)}{z^2} \biggr] f^{L/R}_{\tau, n}(z) = 
M_{n\tau}^2 \, f^{L/R}_{\tau, n}(z),
\en
where
\eq
f^L_{\tau, n}(z) &=& \sqrt{\frac{2\Gamma(n+1)}{\Gamma(n+\tau)}} 
\ \kappa^{\tau}
\ z^{\tau-1/2} \ e^{-\kappa^2 z^2/2} \ L_n^{\tau - 1}(\kappa^2z^2) \,, \\
f^R_{\tau, n}(z) &=& \sqrt{\frac{2\Gamma(n+1)}{\Gamma(n+\tau-1)}} 
\ \kappa^{\tau-1} \ z^{\tau-3/2} \ e^{-\kappa^2 z^2/2} 
\ L_n^{\tau-2}(\kappa^2z^2)
\en
and
\eq
M_{n\tau}^2 = 4 \kappa^2 \Big( n + \tau - 1 \Big)  
\en
with 
\eq\label{norm_cond} 
\int\limits_0^\infty dz \ f^{L/R}_{\tau, n_1}(z) \, 
f^{L/R}_{\tau, n_2}(z) 
\, = \, \delta_{n_1n_2}\; .
\en 
Here 
\eq 
L_n^\tau(x) = \frac{x^{-\tau} e^x}{n!} 
\, \frac{d^n}{dx^n} \Big( e^{-x} x^{\tau+n} \Big)
\en 
are the generalized Laguerre polynomials. 
In the above formulas we substituted $m = \tau  -  3/2$. 

One can see that the functions 
$F^{L/R}_{\tau, n}(z) = e^{- 2 A(z)} \, f^{L/R}_{\tau, n}(z)$ 
have the correct scaling behavior for small $z$
\eq
F^L_{\tau, n}(z) \sim z^{\tau+3/2}\,, \quad\quad
F^R_{\tau, n}(z) \sim z^{\tau+1/2}\,,
\en 
when identified with the corresponding nucleon wave functions with
twist $\tau$
and vanish at large $z$ (confinement). In Ref.~\cite{Gutsche:2012bp} 
it was explicitly demonstrated that the nucleon electromagnetic 
form factors have the correct scaling dependence at large $Q^2$. 

Now we define the 5D fields $\psi^N_{\pm,\tau}(x,z)$ and
$\psi^{\cal R}_{\pm,\tau}(x,z)$, which are holographic analogues 
of the nucleon and Roper resonance, respectively: 
\eq 
\psi^N_{\pm,\tau}(x,z) &=& \frac{1}{\sqrt{2}} \, 
\left[ 
      \psi^{L}_0(x) \ F^{L/R}_{\tau, 0}(z) 
\pm   \psi^{R}_0(x) \ F^{R/L}_{\tau, 0}(z)\right]\,, \nonumber\\
\psi^{\cal R}_{\pm,\tau}(x,z) &=& \frac{1}{\sqrt{2}} \, 
\left[ 
      \psi^{L}_1(x) \ F^{L/R}_{\tau, 1}(z) 
\pm   \psi^{R}_1(x) \ F^{R/L}_{\tau, 1}(z)\right]\,. 
\en
Here we identify the nucleon as the ground state with $n=0$ 
and the Roper resonance as the first radially excited state with $n=1$. 
We should stress again that 5D AdS fields corresponding to the nucleon 
and Roper are products of 4D spinor fields with spin $1/2$ 
and profiles depending on the holographic (scale) variable. 
The free actions of the nucleon and Roper resonance with fixed twist 
dimension $\tau$, are constructed in terms of $\psi^N_{\pm,\tau}(x,z)$ 
and $\psi^{\cal R}_{\pm,\tau}(x,z)$ as: 
\eq 
S_\tau^B = \int d^4x dz \, e^{4A(z)} \, \sum\limits_{i=+,-} 
\, \bar\psi^B_{i,\tau}(x,z) \, 
\biggl\{ i\not\!\partial + \gamma^5\partial_z
+ 2 A^\prime(z) \gamma^5
-  \delta_i \frac{e^{A(z)}}{R} \Big(m + \varphi(z)\Big)  \biggr\} 
\, \psi^B_{i, \tau}(x,z)\,, 
\en 
where $B = N, {\cal R}$. 
In order to take into account higher Fock states in both the nucleon and 
Roper we sum the 5D action $S_\tau^B$ over $\tau$ with adjustable 
coefficients $c_\tau^B$: 
\eq 
S^B = \sum\limits_\tau \, c_\tau^B \, S_\tau^B \,. 
\en 
In Ref.~\cite{Gutsche:2012bp} we 
showed that the $c_\tau^B$ are constrained by the condition  
$\sum_\tau \, c_\tau^B = 1$ in order to get 
the correct normalization of the kinetic term 
$\bar\psi_n(x) i\!\!\not\!\partial\psi_n(x)$ 
of the four-dimensional spinor field. Also this condition is 
consistent with electromagnetic gauge invariance. 

The nucleon and Roper masses are identified with the 
expressions~\cite{Gutsche:2012bp} 
\eq 
M_N &=& 2 \kappa \sum\limits_\tau \, c_\tau^N \, \sqrt{\tau - 1} 
\,, \nonumber\\
M_{\cal R} &=& 2 \kappa \sum\limits_\tau \, c_\tau^{\cal R} \, \sqrt{\tau} \; .
\en  
Integration over the holographic coordinate $z$, with the use of the 
normalization condition (\ref{norm_cond}) for the profile functions 
$f^{L/R}_{\tau, n}(z)$, gives four-dimensional actions for the 
fermion field $\psi_n(x) = \psi^L_n(x) + \psi^R_n(x)$ with 
$n=0$ (for nucleon) and $n=1$ (for Roper): 
\eq
S^B_{4D} \ = \ \int d^4x \, 
\bar\psi_0(x) \biggl[ i \not\!\partial - M_{N} \biggr] \psi_0(x) 
\ + \ \int d^4x \, 
\bar\psi_1(x) \biggl[ i \not\!\partial - M_{\cal R} \biggr] \psi_1(x) 
\,.
\en 
This last equation is a manifestation of the gauge-gravity duality.
It explicitly demonstrates
that effective actions for conventional hadrons in four dimensions
can be generated from actions for bulk
fields propagating in five-dimensional AdS space. The effect of the
extra dimension is encoded in the baryon mass $M_B$. 

In the following we restrict ourselves to the contribution of 
Fock states in both the nucleon and the Roper resonance with 
twist $\tau=3, 4$, and $5$.  
The nucleon mass was already calculated in Ref.~\cite{Gutsche:2012bp}. 
Taking the following choice of parameters $\kappa$, $c_3^N$, and $c_4^N$: 
\eq 
\kappa = 383 \ \mathrm{MeV}\,, \quad c_3^N = 1.25\,, \quad c_4^N = 0.16\,, 
\en   
we reproduce the data for the nucleon (proton) mass  
$M_N^{\rm exp} = 938.27$ MeV. 
Notice that the parameter $c_5^N$ depends on $c_3^N$ and $c_4^N$ as 
\eq 
c_5^N = 1 - c_3^N - c_4^N = - 0.41 \,. 
\en 
Taking the same value of the universal scale parameter 
$\kappa = 383$ MeV we reproduce the world average for the Roper mass 
$M_{\cal R}^{\rm exp} = 1440$ MeV with
\eq 
c_3^{\cal R} = 0.78\,, \quad  c_4^{\cal R} = - 0.16\,, \quad 
c_5^{\cal R} = 1 - c_3^{\cal R} - c_4^{\cal R} = 0.38 \,.  
\en 
One can see that, as in the nucleon case, the $3q$ Fock component gives the 
main contribution to the Roper mass. 

We would like to stress again that the quantities $c_\tau^B$ $(B=N, {\cal R})$ 
are free parameters constrained by the condition 
$\sum_\tau \, c_\tau^B$ = 1. Inclusion of the states for $\tau=6$ 
does not change qualitatively the description of data. On the other hand, 
from the analysis of data on nucleon form factors, nucleon, and Roper mass,  
we found that the contribution of twist-4 Fock states (containing 
three quarks and one gluon) is always suppressed in comparison 
with leading $3q$ component and twist-5 Fock states containing 
a sizeable $3q+q\bar q$ component.  

Next we will study the role of different Fock components 
in the electroproduction properties of the Roper resonance.  

\section{Electroproduction of the Roper resonance 
$N + \gamma^\ast \to {\cal R}$} 

\subsection{Kinematics} 

The electromagnetic transition between the nucleon and the Roper resonance, 
due to Lorentz and gauge invariance, is defined by the following 
matrix element: 
\eq 
M^\mu(p_1,\lambda_1;p_2,\lambda_2) = \bar u_{\cal R}(p_1,\lambda_1) 
\left[ \gamma^\mu_\perp \, F_1(q^2) \, + \, 
i \sigma^{\mu\nu} \frac{q_\nu}{M_{\cal R}} \, F_2(q^2) \, \right] 
u_N(p_2,\lambda_2)\,, 
\quad \gamma^\mu_\perp 
= \gamma^\mu - q^\mu \frac{\not\! q}{q^2}\,, 
\en 
which obeys current conservation
\eq 
q_\mu \, M^\mu \, = \, 0\,,
\en 
where $(p_1, \lambda_1)$, $\ (p_2, \lambda_2)$, $\ (q = p_1 - p_2, \, 
\lambda = \lambda_1 + \lambda_2)$ are the (momenta, helicity) 
of the Roper resonance, the nucleon and the photon, respectively. 
We shall work in the rest frame of the daughter baryon (Roper) with 
the parent baryon (nucleon) moving in the negative $z$ direction 
($z$ axis is directed along the photon 3-momentum):
\eq 
p_1^\mu = (M_{\cal R}, \vec{\bf 0})\,, \quad 
p_2^\mu = (E, 0, 0, -|{\bf p}|)\,, \quad  
q^\mu   = (q_0, 0, 0, |{\bf p}|)\,, 
\en 
where 
\eq 
E =  \frac{Q_+}{2M_{\cal R}} - M_N\,, \quad 
|{\bf p}| = \frac{\sqrt{Q_+ Q_-}}{2M_{\cal R}}\,, \quad 
Q_\pm = M_\pm^2 + Q^2\,, \quad 
M_\pm = M_{\cal R} \pm M_N \,, \quad 
Q^2 = - q^2 \,. 
\en 
Alternative sets of transition form factors can be found in 
Refs.~\cite{Weber:1989fv,Aznauryan:2007ja,Tiator:2008kd}. 

Now we introduce the helicity amplitudes 
$H_{\lambda_2\lambda}$, which in turn can be 
related to the invariant form factors $F_i$ (see details in 
Refs.~\cite{Kadeer:2005aq,Faessler:2009xn,Branz:2010pq}). 
The pertinent relation is
\eq
H_{\lambda_2\lambda} = M_\mu(p_1,\lambda_1;p_2,\lambda_2)
\, \epsilon^{\ast \, \mu}(q,\lambda) \,, 
\en 
where the polarization vectors of the outgoing photon 
$\epsilon^{\ast \,\mu}(q, \lambda)$ are written as 
\eq 
\epsilon^{\ast \,\mu}(q,\pm 1) 
= \frac{1}{\sqrt{2}} (0, \mp 1, i, 0)\,, \quad 
\epsilon^{\ast \, \mu}(q,0) 
= \frac{1}{\sqrt{Q^2}} (|{\bf p}|, 0, 0, q_0) \,.   
\en 
The $J=\frac{1}{2}$ baryon spinors are 
given by 
\eq 
\bar u_{\cal R}\Big(p_1, \pm \frac{1}{2}\Big) &=& \sqrt{2M_{\cal R}} \, 
\Big( \chi_\pm^\dagger, 0 \Big)\,, \nonumber\\
u_N\Big(p_2, \mp \frac{1}{2}\Big) &=& 
\sqrt{E + M_N} \, 
\left(
\begin{array}{c}
\chi_\pm \\
\frac{\mp |{\bf p}|}{E + M_N} \, \chi_\pm \\
\\
\end{array}
\right)
\en 
Here $\chi_+ = \left(
\begin{array}{l}
1 \\
0 \\
\end{array} \right)$  
and $\chi_- = \left(
\begin{array}{l}
0 \\
1 \\
\end{array} \right)$ 
are two-component Pauli spinors. 

After straightforward 
calculations~\cite{Kadeer:2005aq,Faessler:2009xn,Branz:2010pq}  
we find  
\eq 
H_{\pm\frac{1}{2}0} &=& \sqrt{\frac{Q_-}{Q^2}} \, 
\left( 
F_1 M_+  - F_2 \frac{Q^2}{M_{\cal R}} \right) \,, \\
H_{\pm\frac{1}{2}\pm 1} &=& - \sqrt{2 Q_-} \, 
\left( F_1 + F_2 \frac{M_+}{M_{\cal R}} \right) \,. 
\en 
The alternative set of the helicity amplitudes $(A_{1/2}, S_{1/2})$ 
is related to the set $(H_{\frac{1}{2}0},H_{\frac{1}{2}1})$ 
as~\cite{Weber:1989fv,Capstick:1994ne,Copley:1972tu,%
Aznauryan:2007ja,Tiator:2008kd} 
\eq 
A_{1/2} = - b \, H_{\frac{1}{2}1}\,, \quad 
S_{1/2} = b \, \frac{|{\bf p}|}{\sqrt{Q^2}} \, H_{\frac{1}{2}0} 
\en 
where 
\eq 
b = \sqrt{\frac{\pi\alpha}{M_+ M_- M_N }} 
\en 
and $\alpha = 1/137.036$ is the fine-structure constant.  

\subsection{Form factors of the $N + \gamma \to {\cal R}$ 
transition} 

In our approach the matrix element describing the $N + \gamma \to {\cal R}$ 
transition is generated by the following 5d action:  
\eq 
S_{\rm int} =  \int d^4x dz \, \sqrt{g} \, e^{-\varphi(z)}
{\cal L}_{\rm int}(x,z)\; .
\en 
${\cal L}_{\rm int}(x,z)$ is the interaction Lagrangian of 
two fermion fields (holographically corresponding to the nucleon and 
Roper resonance) and vector field (holographically corresponding to 
the electromagnetic field): 
\eq 
{\cal L}_{\rm int}(x,z) &=& 
\sum\limits_{i=+,-; \,\tau} \, c_\tau^{{\cal R}N} \, 
\bar\psi_{i,\tau}^{\cal R}(x,z) \, \hat{\cal V}_i(x,z) \, 
\psi_{i,\tau}^N(x,z) \, + \, {\rm H.c.}\,, \nonumber\\ 
\hat{\cal V}_\pm(x,z)  &=&  \tau_3 \Gamma^M  V_M(x,z) \, \pm \, 
\frac{i}{4} \, \eta_V \,  [\Gamma^M, \Gamma^N] \, V_{MN}(x,z)  
\, \pm \, g_V\, \tau_3 \, \Gamma^M \, i\Gamma^z \, V_M(x,z)  \,, 
\en  
where $c_\tau^{{\cal R}N}$ is the set of parameters mixing the contribution 
of AdS fermion fields with different twist dimension. Here
$\eta_V = {\rm diag}(\eta_p,\eta_n)$ and 
$\tau_3$ is the Pauli isospin matrix, $\eta_p$, $\eta_n$, $g_V$ 
are the coupling constants, $V_M(x,z)$ is the AdS vector field, and 
$V_{MN} = \partial_M V_N - \partial_N V_M$ is its stress tensor. 

The expressions for the $F_1$ and $F_2$ form factors 
are given by: 
\eq 
F_1^p(Q^2) &=& C_1(Q^2) + g_V C_2(Q^2) 
+ \eta_p C_3(Q^2) \,, \\
F_2^p(Q^2) &=& \eta_p C_4(Q^2) \, 
\en 
for the case of $p + \gamma \to {\cal R}_p$ transition, and 
\eq 
F_1^n(Q^2) &=& - C_1(Q^2) - g_V C_2(Q^2) 
+ \eta_n C_3(Q^2) \,, \\
F_2^n(Q^2) &=& \eta_n C_4(Q^2) \, 
\en 
for the case of $n + \gamma \to {\cal R}_n$ transition. 
Here, ${\cal R}_p$ and ${\cal R}_n$ are the members of the Roper isospin
doublet, and $C_i(Q^2)$ are the structure integrals 
(see explicit expressions in the Appendix): 
\eq 
C_{1}(Q^2) &=& \frac{1}{2} \, \int\limits_0^\infty dz \, V(Q,z) 
\ \sum\limits_\tau \, c_\tau^{{\cal R}N} \, 
\biggl( f^L_{\tau,0}(z) f^L_{\tau,1}(z) + 
f^R_{\tau,0}(z) f^R_{\tau,1}(z) \biggr)\,, 
\nonumber\\ 
C_{2}(Q^2) &=& \frac{1}{2} \, \int\limits_0^\infty dz \, 
V(Q,z) \ \sum\limits_\tau \, c_\tau^{{\cal R}N} \, 
\biggl( f^R_{\tau,0}(z) f^R_{\tau,1}(z) - 
f^L_{\tau,0}(z) f^L_{\tau,1}(z) \biggr)\,, 
\nonumber\\ 
C_{3}(Q^2) &=& \frac{1}{2} \, \int\limits_0^\infty dz z \, \partial_z V(Q,z) 
\ \sum\limits_\tau \, c_\tau^{{\cal R}N} \, 
\biggl( f^L_{\tau,0}(z) f^L_{\tau,1}(z) - 
f^R_{\tau,0}(z) f^R_{\tau,1}(z) \biggr)\,, 
\nonumber\\ 
C_{4}(Q^2) &=& \frac{M_+}{2} \, \int\limits_0^\infty dz z \, V(Q,z) 
\ \sum\limits_\tau \, c_\tau^{{\cal R}N} \, 
\biggl( f^L_{\tau,0}(z) f^R_{\tau,1}(z) +  
f^L_{\tau,1}(z) f^R_{\tau,0}(z) \biggr) 
\,. \label{Ci} 
\en 
The functions $f^{R/L}_{\tau,n=0,1}(z)$ 
are the bulk profiles of fermions with specific $n=0$ or $1$. 
$V(Q,z)$ is the bulk-to-boundary propagator of 
the transverse massless vector bulk field (the holographic analogue of 
the electromagnetic field), defined as 
\eq
V_\mu(x,z) = \int \frac{d^4 q}{(2\pi)^4} e^{-iqx} V_\mu(q) V(q,z), 
\en
which obeys the following EOM 
\eq
  \partial_z \biggl( \frac{e^{-\varphi(z)}}{z} \partial_z V(q,z) \biggr)
+ q^2 \, \frac{e^{-\varphi(z)}}{z} V(q,z) = 0 \,. 
\en 
In the soft-wall model the solution for $V(Q,z)$ is given in analytical  
form in terms of the gamma $\Gamma(n)$ and Tricomi $U(a,b,z)$ functions:  
\eq\label{V} 
V(Q,z) = \Gamma\biggl(1 + \frac{Q^{2}}{4 \kappa^{2}}\biggr) 
U\biggl(\frac{Q^{2}}{4 \kappa^{2}}, 0, \kappa^2 z^2\biggr)\;.
\en 
The bulk-to-boundary propagator $V(Q,z)$ obeys the normalization
condition $V(0,z) = 1$, consistent with 
gauge invariance, and fulfills the following ultraviolet (UV) and 
infrared (IR) boundary conditions : 
\eq 
V(Q,0) = 1\,, \quad\quad V(Q,\infty) = 0 \,. 
\en 
The UV boundary condition corresponds to the local (structureless) 
coupling of the electromagnetic field to matter fields, while 
the IR boundary condition implies that the vector field
vanishes at $z=\infty$. 

In order to obtain analytical expressions for the functions $C_i(Q^2)$, 
it is convenient to use the integral representation for $V(Q,z)$ 
introduced in Ref.~\cite{Soft_wall4a} 
\eq 
\label{VInt}
V (Q,z) = \kappa^{2} z^{2} \int_{0}^{1} \frac{dx}{(1-x)^{2}} 
\, x^{\frac{Q^{2}}{4 \kappa^{2}}} \, 
e^{- \displaystyle{\frac{\kappa^2 z^2 x}{1-x} }}\,,   
\en 
where the variable $x$ is equivalent to the light-cone momentum 
fraction~\cite{Soft_wall2ab}. 

There are a few very important properties of the $C_i(Q^2)$ functions. 
Namely, at $Q^2 = 0$ they are normalized as  
\eq 
C_1(0) = C_2(0) = C_3(0) = 0\,, 
\quad C_4(0) = - \frac{M_+}{2} \sum\limits_\tau \, c_\tau^{{\cal R}N} \,. 
\en 
The normalizations of $C_i$ $(i=1,2,3)$ are consistent with 
gauge invariance. 

The analytical power scaling of the $C_i(Q^2)$ functions 
and therefore of the form factors $F_i^N(Q^2)$, 
defining the electromagnetic transition between nucleon and Roper resonance, 
is in accordance with quark-counting rules for large momentum transfers.
In particular, the leading $3q$ contribution in $C_i(Q^2)$ and $F_i^N(Q^2)$ 
scales for $Q^2 \to \infty$ as 
\eq 
C_1 (Q^2) \sim C_2 (Q^2) \sim C_3(Q^2) \sim \frac{1}{Q^4}\,, 
\quad 
C_4(Q^2) \sim \frac{1}{Q^6}\, 
\en 
and 
\eq 
F_1^N(Q^2) \sim \frac{1}{Q^4}\,,  
\quad 
F_2^N(Q^2) \sim \frac{1}{Q^6}\, 
\en
for the case of $p + \gamma \to {\cal R}_p$ transition, and similarly for
\eq
F_1^n(Q^2) &=& - C_1(Q^2) - g_V C_2(Q^2)
+ \eta_n C_3(Q^2) \,, \\
F_2^n(Q^2) &=& \eta_n C_4(Q^2) \, .
\en

Following Ref.~\cite{Tiator:2008kd} we define the transition charge 
density for the unpolarized $N \to R$ transition:  
\eq 
\rho_0(\vec{b}_\perp\,) = \int\frac{d^2\vec{q}_\perp}{(2\pi)^2} \,
e^{-i\vec{q}_\perp \vec{b}_\perp} \
\frac{1}{2P^+} \, \la P^+, \frac{\vec{q}_\perp}{2}, \lambda | J^+(0) |
P^+, -\frac{\vec{q}_\perp}{2}, \lambda\ra  
\en 
and for the transversely polarized nucleon and Roper resonance, 
both along the direction of 
$\vec{S}_\perp = \cos\phi_S \hat{e}_x + \sin\phi_S \hat{e}_y$:  
\eq 
\rho_T(\vec{b}_\perp\,) = \int\frac{d^2\vec{q}_\perp}{(2\pi)^2} \,
e^{-i\vec{q}_\perp \vec{b}_\perp} \
\frac{1}{2P^+} \, \la P^+, \frac{\vec{q}_\perp}{2}, s_\perp | J^+(0) |
P^+, -\frac{\vec{q}_\perp}{2}, s_\perp \ra 
\en 
where $\vec{b}_\perp$ is the position in the $(xy)$ plane from transverse
c.m. of the baryons and $s_\perp$ is the nucleon spin projection along 
the direction of $\vec{S}_\perp$. 

\section{Results} 

In this section we present the numerical analysis of the 
physical observables of the  electromagnetic 
nucleon-Roper transition: 
form factors, helicity amplitudes and transition charge radii. 
As in the case of the nucleon electromagnetic form factors, 
for the electroproduction of the Roper resonance the main contribution is 
given by the leading $3q$ component. The higher Fock components 
compensate each other. We illustrate this feature in 
Figs. 1--4. In particular, the plots of the leading ($3q$ Fock component) 
and full results (including $3q$, 4 and 5 partonic contributions) 
practically coincide. 
In Table I we present full results and leading $3q$ contributions 
(in brackets) for the helicity amplitudes $A^N_{1/2}(0)$ and 
$S^N_{1/2}(0)$, $N=p, n$. 

Finally, the free parameters are fixed as: 
\eq 
& &
g_V = 1\,, \hspace*{.5cm} 
   \eta_p =  0.453\,, \hspace*{.5cm} 
   \eta_n = -0.279\,, \nonumber\\
& &
c_3^{{\cal R}N} =  0.72\,, \hspace*{.5cm}
c_4^{{\cal R}N} = - c_5^{{\cal R}N} = -0.18\,. 
\en 
From Figs. 5 and 6 it should be evident that our results for 
the helicity amplitudes in the proton case have qualitative agreement 
with the present data of the CLAS Collaboration~\cite{Aznauryan:2009mx}. 
Within the current approach it is difficult to reproduce the maximum  
of data for $A^p_{1/2}$ at about 2 GeV$^2$. Higher-twist contributions 
cannot improve this situation as discussed before. We recently proposed an 
extension of our soft-wall model including a longitudinal wave function 
in the case of mesons. In the future we plan to do a similar extension in 
the baryon sector, which can help to improve the fit in the 
intermediate $Q^2$ region.  
Further data for the helicity amplitudes in the region 
from 1.6 to 4 GeV$^2$ could be accumulated at the upgraded facilities 
of JLab and certainly help to clarify the theoretical understanding. 
In Figs.7-10 we show our predictions for the proton- and neutron-case
helicities up to 12 GeV$^2$. For completeness in Figs.11-20 we plot the
2D- and 3D-images of the charge distributions.

\section{Conclusion} 

We presented a detailed analysis of the electroproduction of the 
Roper resonance in the framework of the soft-wall AdS/QCD model. 
We showed that the Roper mass is mainly generated by the contribution 
of the leading $3q$ Fock states, and contributions of 
the higher (4- and 5-partonic) Fock states are small (a similar 
result was found for the nucleon~\cite{Gutsche:2012bp}). 
In the case of the electroproduction amplitude the leading 
$3q$ Fock state plays the dominant role and higher Fock states 
compensate each other. We hope that our predictions will be 
useful for the JLab experiments and for theoretical 
investigations on the nature of the Roper resonance. 

In the future we plan to apply this formalism to other baryon resonances with 
adjustable quantum numbers $n$, $J$, and $L$. We suppose that the 
parameters related to the mixing of the different Fock components 
are not necessarily the same for all baryons. Therefore, the 
set of parameters $c_\tau^B$ defining the mixing of the Fock components 
in the specific baryon state and of parameters $c_\tau^{B_1B_2}$ defining 
the coupling of two Fock components of twist $\tau$ in baryons $B_1$ and 
$B_2$ could be varied. 

\begin{acknowledgments}

This work was supported by the DFG under Contract No. LY 114/2-1,
by the Federal Targeted Program ``Scientific
and scientific-pedagogical personnel of innovative Russia''
Contract No. 02.740.11.0238, by FONDECYT (Chile) under Grant No. 1100287 
and by CONICYT (Chile) under Grant No. 7912010025. 
V. E. L. would like to thank Departamento de F\'\i sica y Centro
Cient\'\i fico Tecnol\'ogico de Valpara\'\i so (CCTVal), Universidad
T\'ecnica Federico Santa Mar\'\i a, Valpara\'\i so, Chile for warm
hospitality.

\end{acknowledgments}

\appendix

\section{Structure integrals  $C_i(Q^2)$} 

The structure integrals 
$C_i(Q^2)$ are given by the expressions: 
\eq 
C_{i}(Q^2) &=& \sum\limits_\tau \, c_\tau^{{\cal R}N} \, 
C_{i}^\tau(Q^2)\,, \\
C_{1}^\tau(Q^2) &=& 
\frac{a}{2} \, B(a+1,\tau+1) \, 
\biggl( \sqrt{\tau-1} \biggl(1 + \frac{a+1}{\tau} \biggr) + \sqrt{\tau}
\biggr) \,, \\
C_{2}^\tau(Q^2) &=& 
\frac{a}{2} \, B(a+1,\tau+1) \, 
\biggl( \sqrt{\tau-1} \biggl(1 + \frac{a+1}{\tau} \biggr) - \sqrt{\tau}
\biggr) \,, \\
C_{3}^\tau(Q^2) &=& \frac{a}{\tau+1} \, 
B(a+1,\tau+2) \,  
\biggl( 
\sqrt{\tau} \, (1 - a\tau) + \sqrt{\tau-1} \, (a(\tau-1) - 1) 
\, \biggl(1 + \frac{a + 2}{\tau} \biggr)\biggr)\,, \\
C_{4}^\tau(Q^2) &=& \frac{M_N+M_{\cal R}}{2} \, B(a+1,\tau+1) \, 
\biggl( 
a(\tau - 1) - \tau - 1 + a \sqrt{\tau (\tau - 1)}  
\biggr) \,, 
\en 
where $a = Q^2/(4\kappa^2)$.

\newpage
\begin{table}
\begin{center}
\caption{Helicity amplitudes 
$A^N_{1/2}(0)$ and $S^N_{1/2}(0)$, $N=p, n$: 
full results and leading 3q contributions (in brackets).} 

\vspace*{.25cm}

\def\arraystretch{1.5}
    \begin{tabular}{|c|c|c|}
      \hline
Quantity & Our results & Data~\cite{PDG} \\
\hline
$A^p_{1/2}(0)$ (GeV$^{-1/2}$)  & -0.065 (-0.065) &  -0.065 $\pm$ 0.004 \\
\hline
$A^n_{1/2}(0)$ (GeV$^{-1/2}$)  &  0.040 (0.040)  &  0.040 $\pm$ 0.010 \\
\hline
$S^p_{1/2}(0)$ (GeV$^{-1/2}$)  &  0.047 (0.048)  &   \\
\hline
$S^n_{1/2}(0)$ (GeV$^{-1/2}$)  & -0.044 (-0.045) &   \\
\hline
\end{tabular}
\end{center}
\end{table} 

\newpage

\begin{figure} 
\begin{center}
\vspace*{1.25cm}
\epsfig{figure=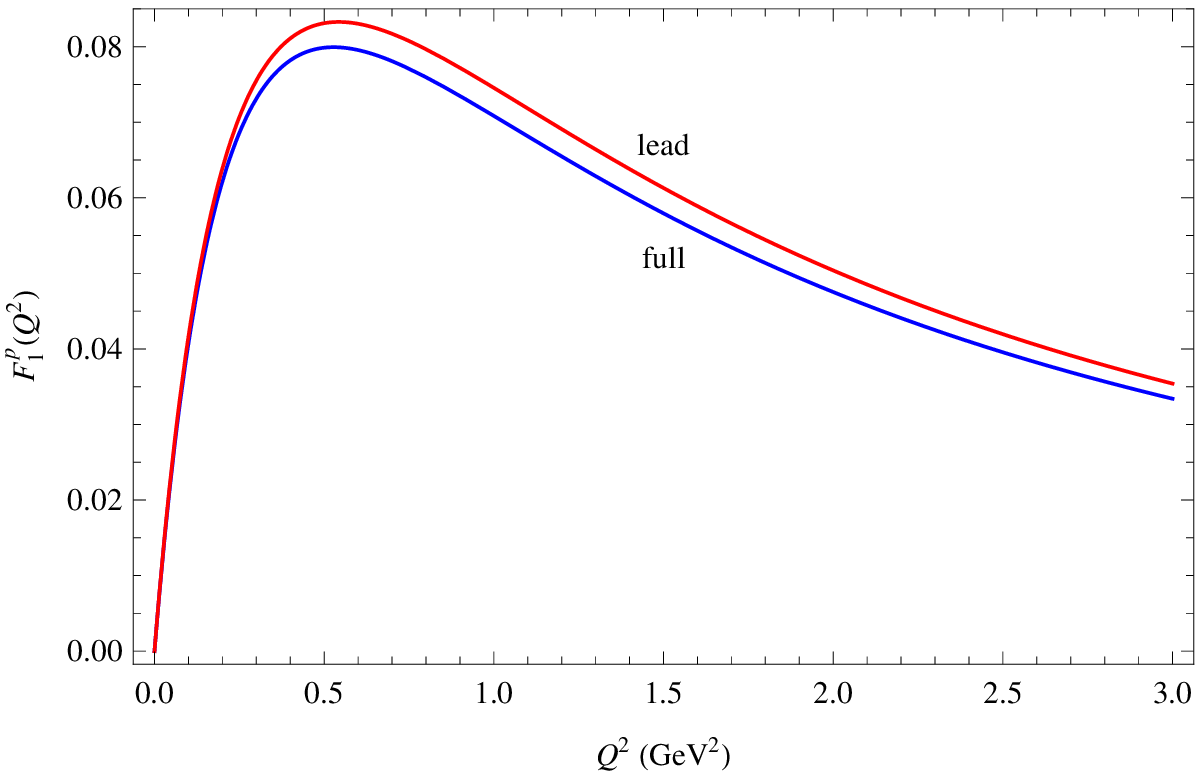,scale=0.6}
\end{center}
{\bf Fig.1:} $F_1^p(Q^2)$ form factor.  

\vspace*{.5cm}

\begin{center}
\vspace*{1.25cm}
\epsfig{figure=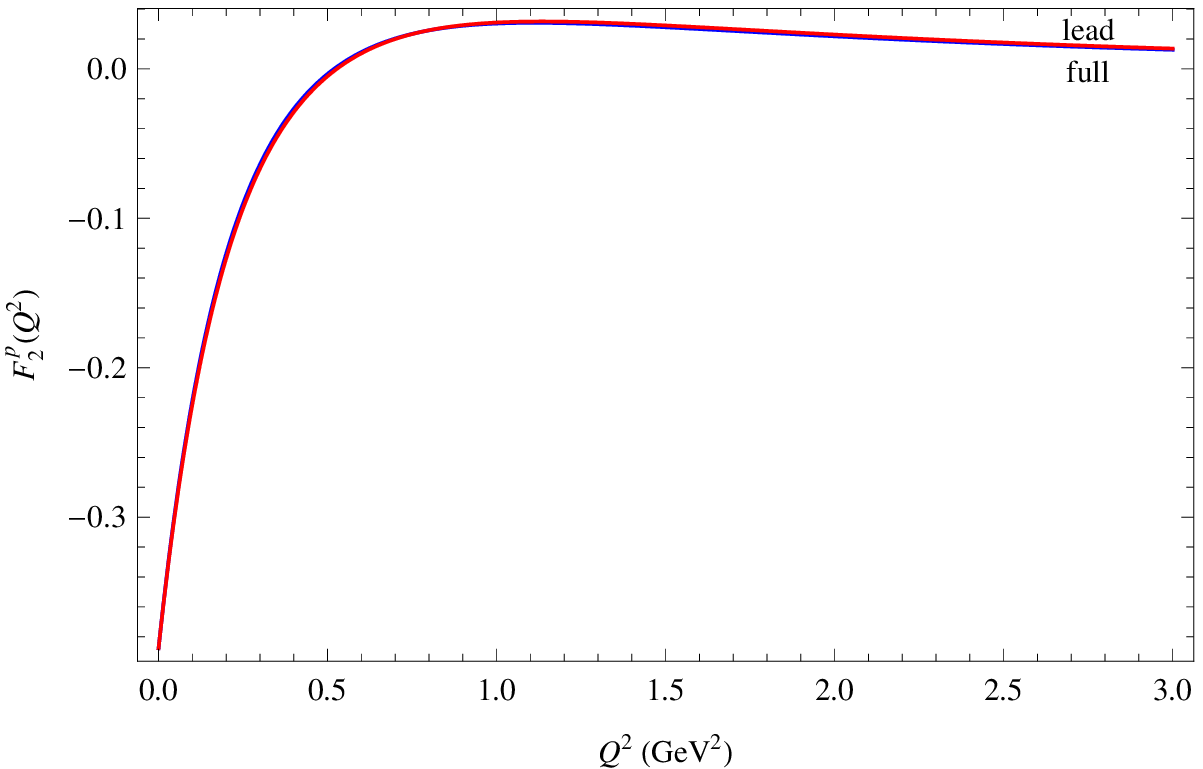,scale=0.6}
\end{center}
{\bf Fig.2:} $F_2^p(Q^2)$ form factor.  
\end{figure} 

\newpage

\begin{figure} 
\begin{center}
\vspace*{1.25cm}
\epsfig{figure=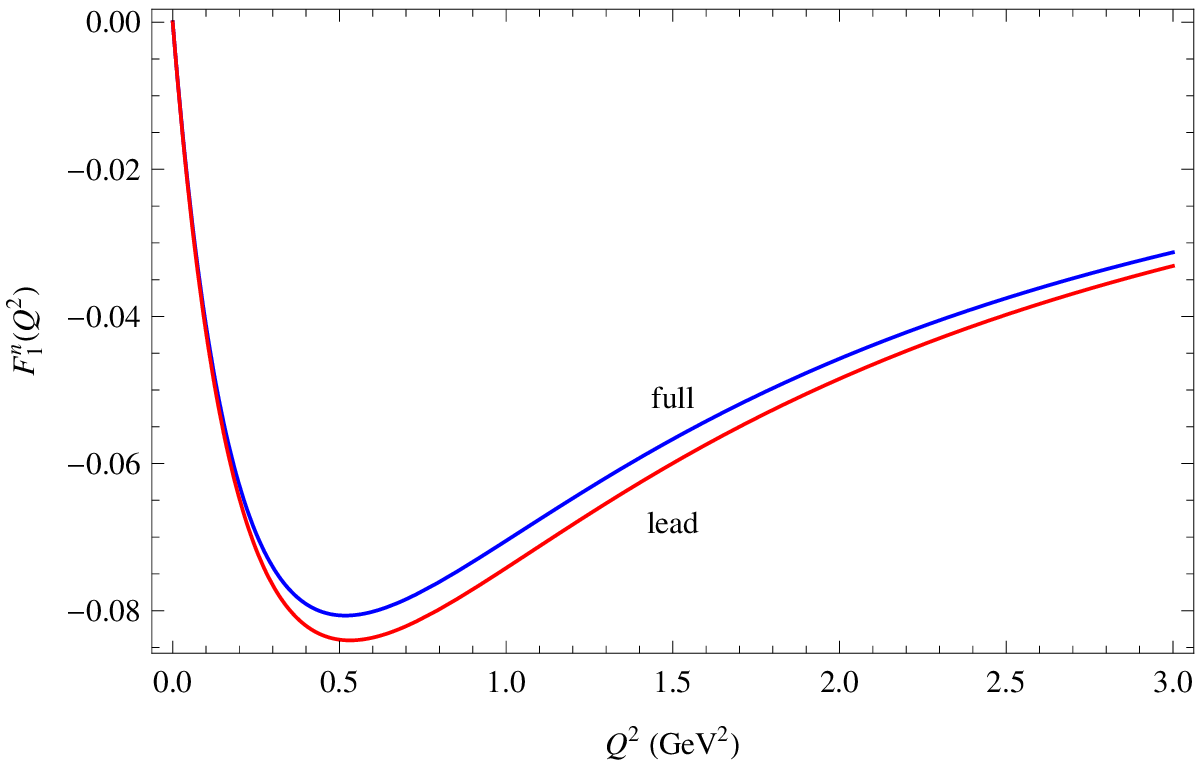,scale=0.6}
\end{center}
{\bf Fig.3:} $F_1^n(Q^2)$ form factor.  

\vspace*{.5cm}

\begin{center}
\vspace*{1.25cm}
\epsfig{figure=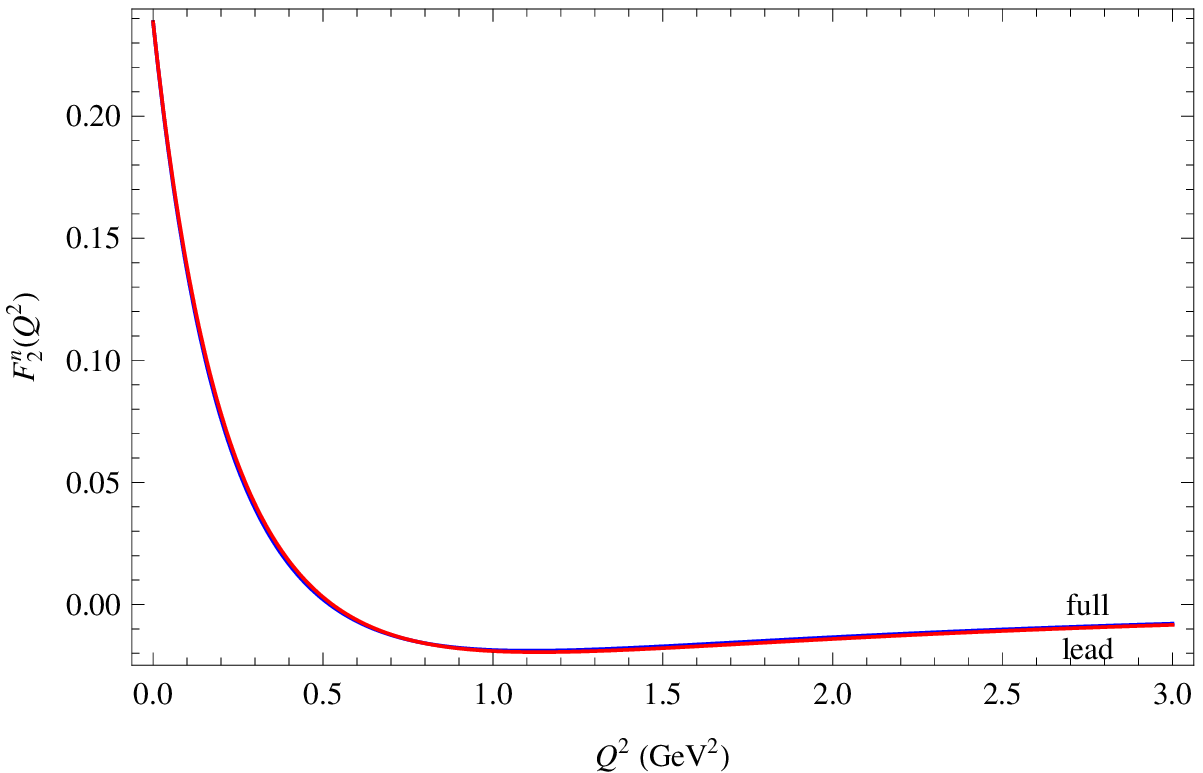,scale=0.6}
\end{center}
{\bf Fig.4:} $F_2^n(Q^2)$ form factor.  
\end{figure} 

\newpage 

\begin{figure} 
\begin{center}
\vspace*{1.25cm}
\epsfig{figure=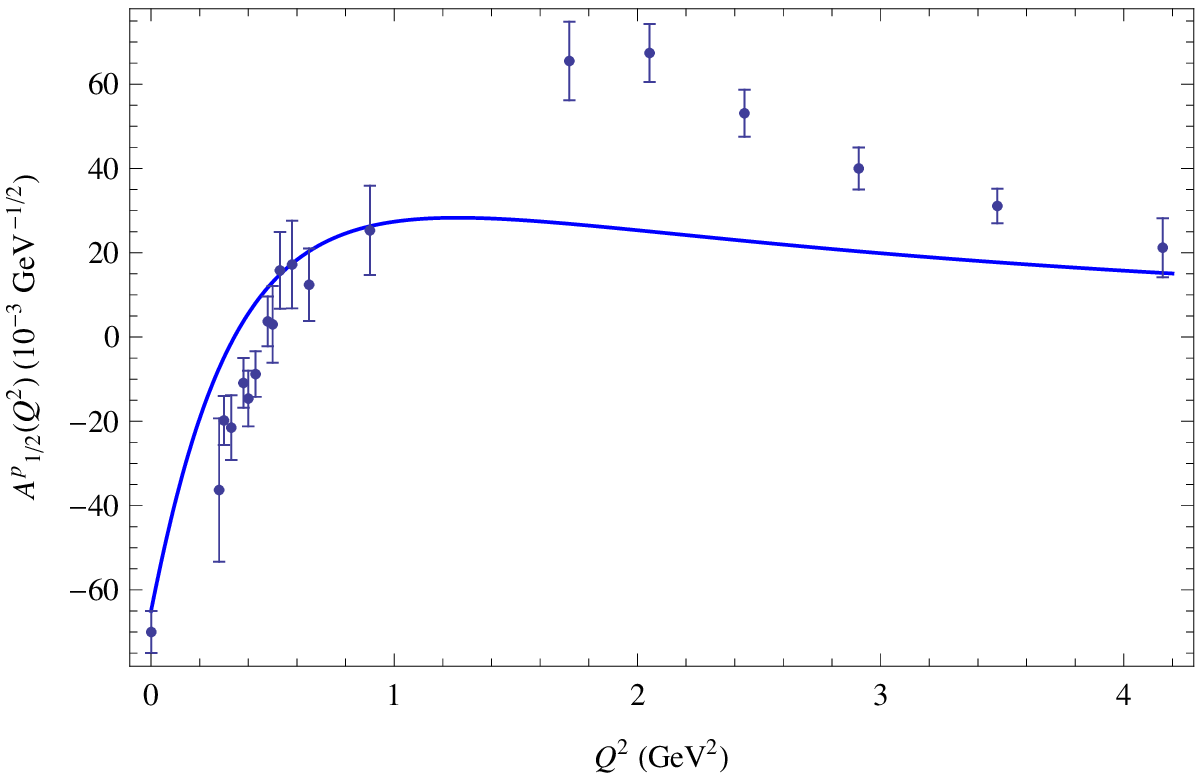,scale=0.6}
\end{center}
{\bf Fig.5:} Helicity amplitude $A^p_{1/2}(Q^2)$ up to $Q^2 = 4$ GeV$^2$.  

\vspace*{.5cm}

\begin{center}
\vspace*{1.25cm}
\epsfig{figure=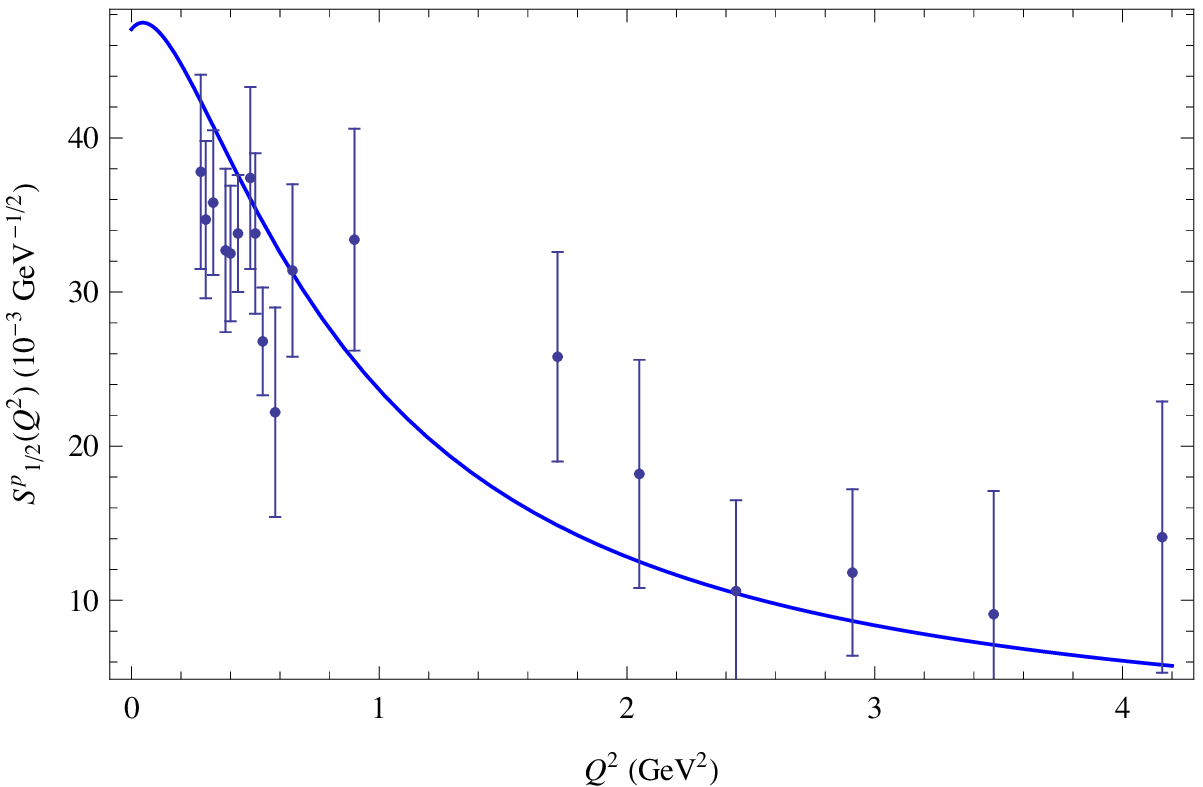,scale=0.6}
\end{center}
{\bf Fig.6:} Helicity amplitude $S^p_{1/2}(Q^2)$ up to $Q^2 = 4$ GeV$^2$.  
\end{figure} 

\newpage 

\begin{figure} 
\begin{center}
\vspace*{1.25cm}
\epsfig{figure=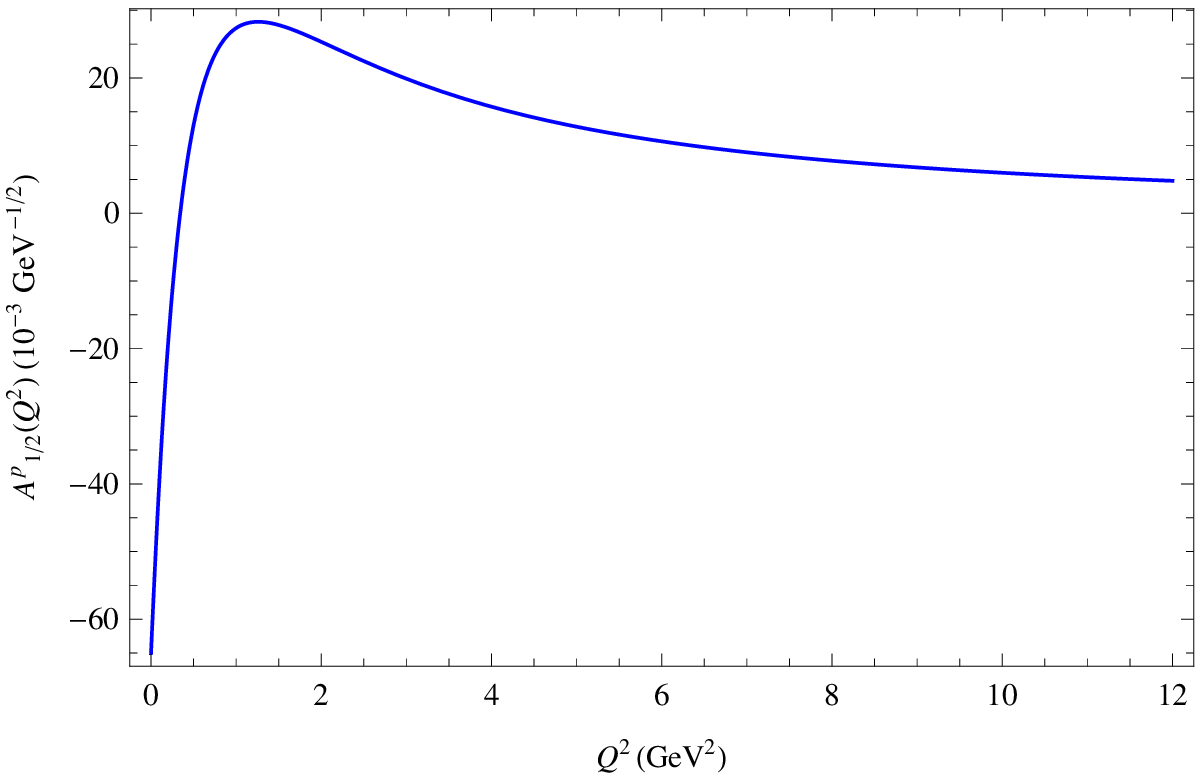,scale=0.6}
\end{center}
{\bf Fig.7:} Helicity amplitude $A^p_{1/2}(Q^2)$ up to $Q^2 = 12$ GeV$^2$.

\vspace*{.5cm}

\begin{center}
\vspace*{1.25cm}
\epsfig{figure=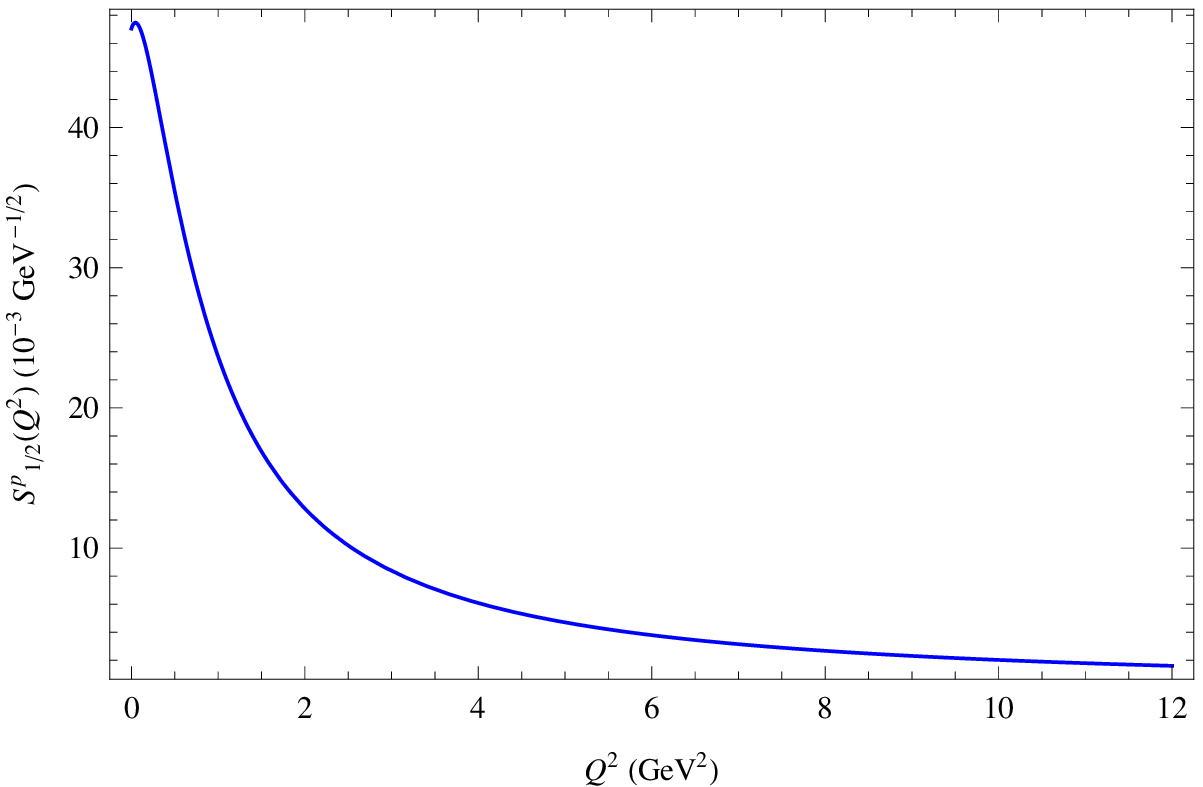,scale=0.6}
\end{center}
{\bf Fig.8:} Helicity amplitude $S^p_{1/2}(Q^2)$ up to $Q^2 = 12$ GeV$^2$.  
\end{figure}

\newpage 

\begin{figure} 
\begin{center}
\vspace*{1.25cm}
\epsfig{figure=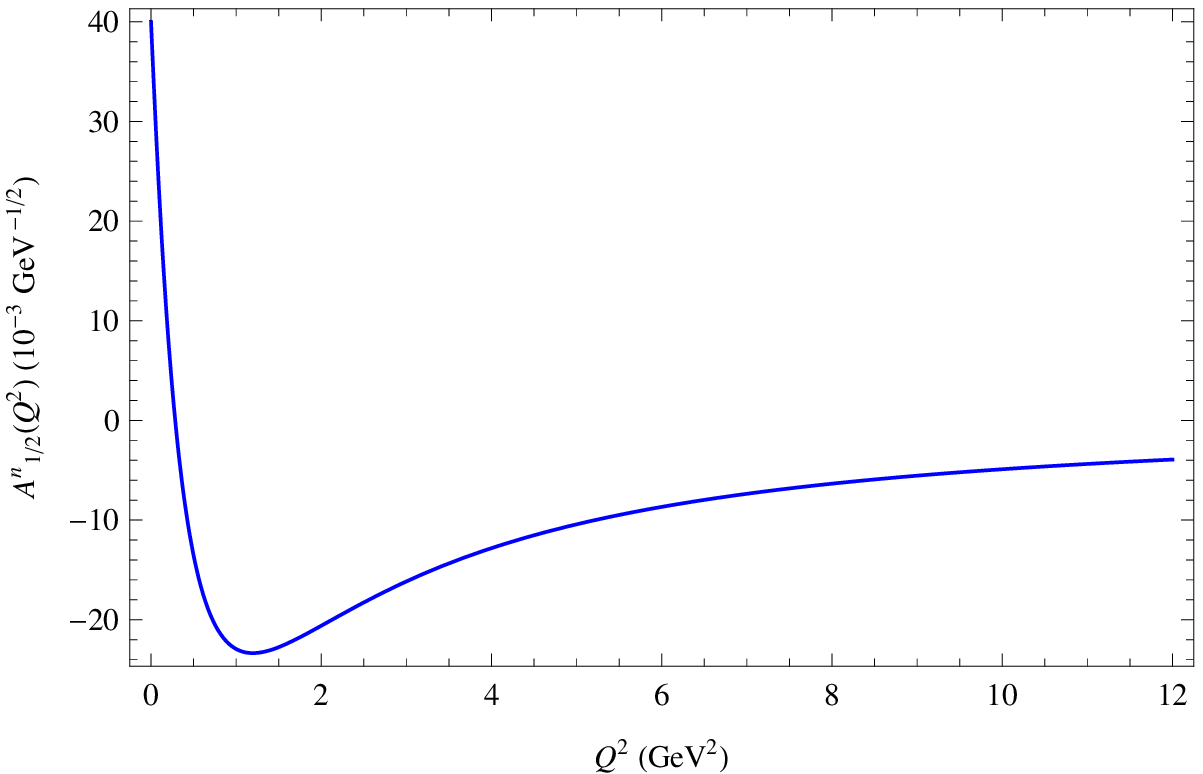,scale=0.6}
\end{center}
{\bf Fig.9:} Helicity amplitude $A^n_{1/2}(Q^2)$ up to $Q^2 = 12$ GeV$^2$.  

\vspace*{.5cm}

\begin{center}
\vspace*{1.25cm}
\epsfig{figure=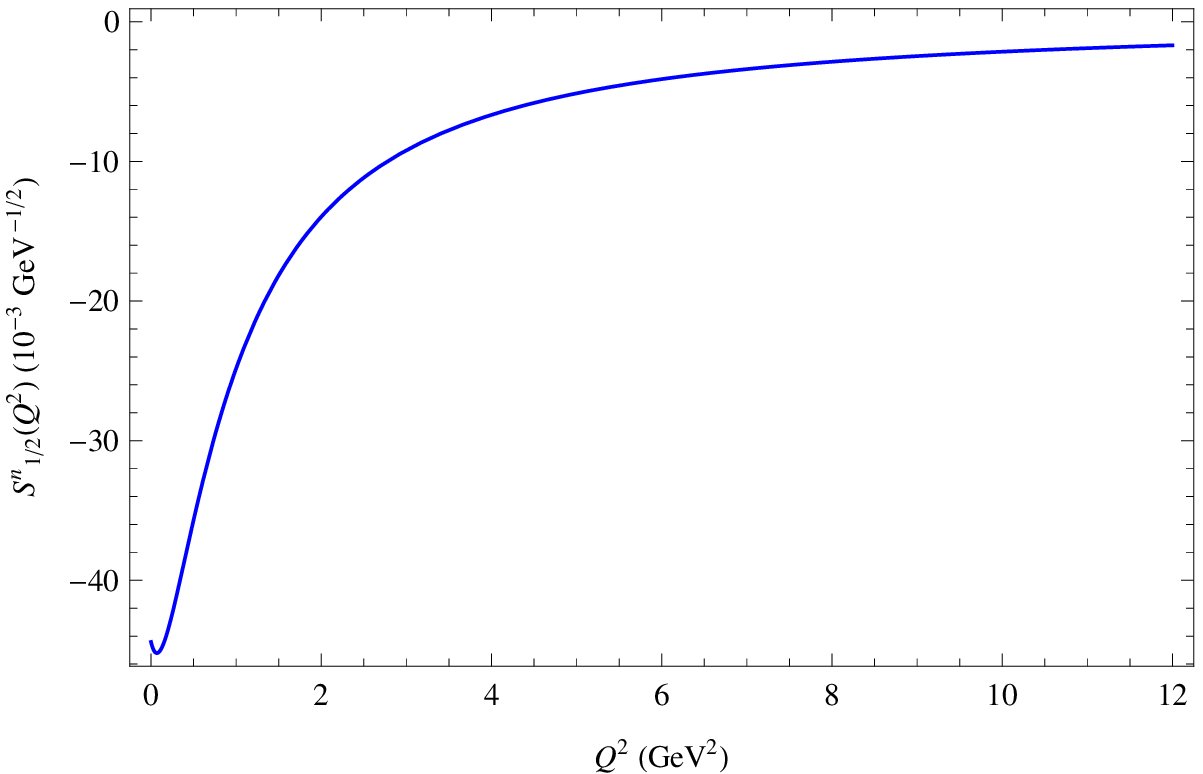,scale=0.6}
\end{center}
{\bf Fig.10:} Helicity amplitude $S^n_{1/2}(Q^2)$ up to $Q^2 = 12$ GeV$^2$.  
\end{figure}

\newpage 

\begin{figure} 
\begin{center}
\vspace*{1.25cm}
\epsfig{figure=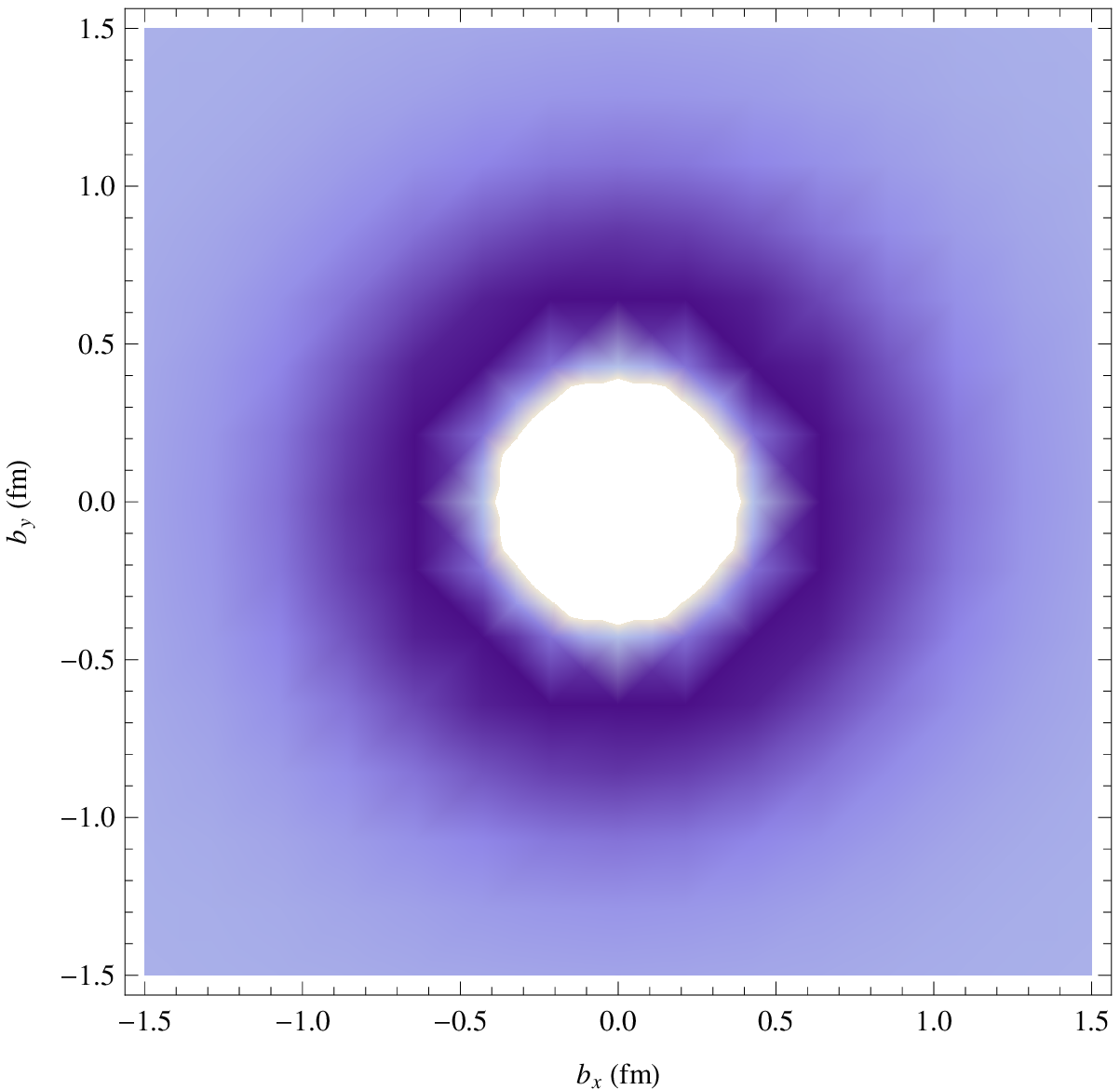,scale=0.6}
\end{center}
{\bf Fig.11:} Charge density $\rho_0^p(b_x,b_y)$. 

\vspace*{.5cm}

\begin{center}
\vspace*{1.25cm}
\epsfig{figure=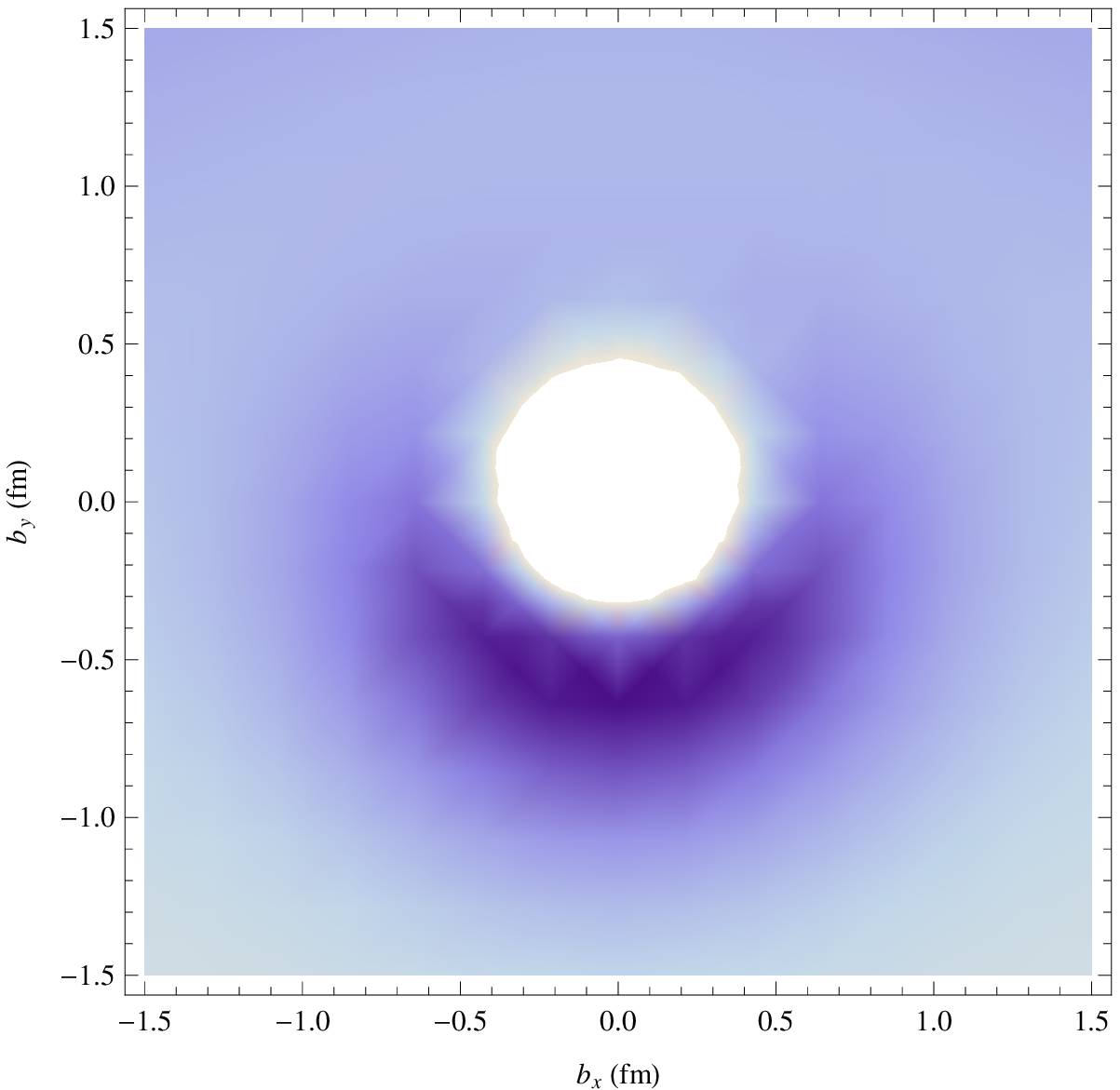,scale=0.6}
\end{center}
{\bf Fig.12:} Charge density $\rho_T^p(b_x,b_y)$. 
\end{figure} 

\newpage 

\begin{figure} 
\begin{center}
\vspace*{1.25cm}
\epsfig{figure=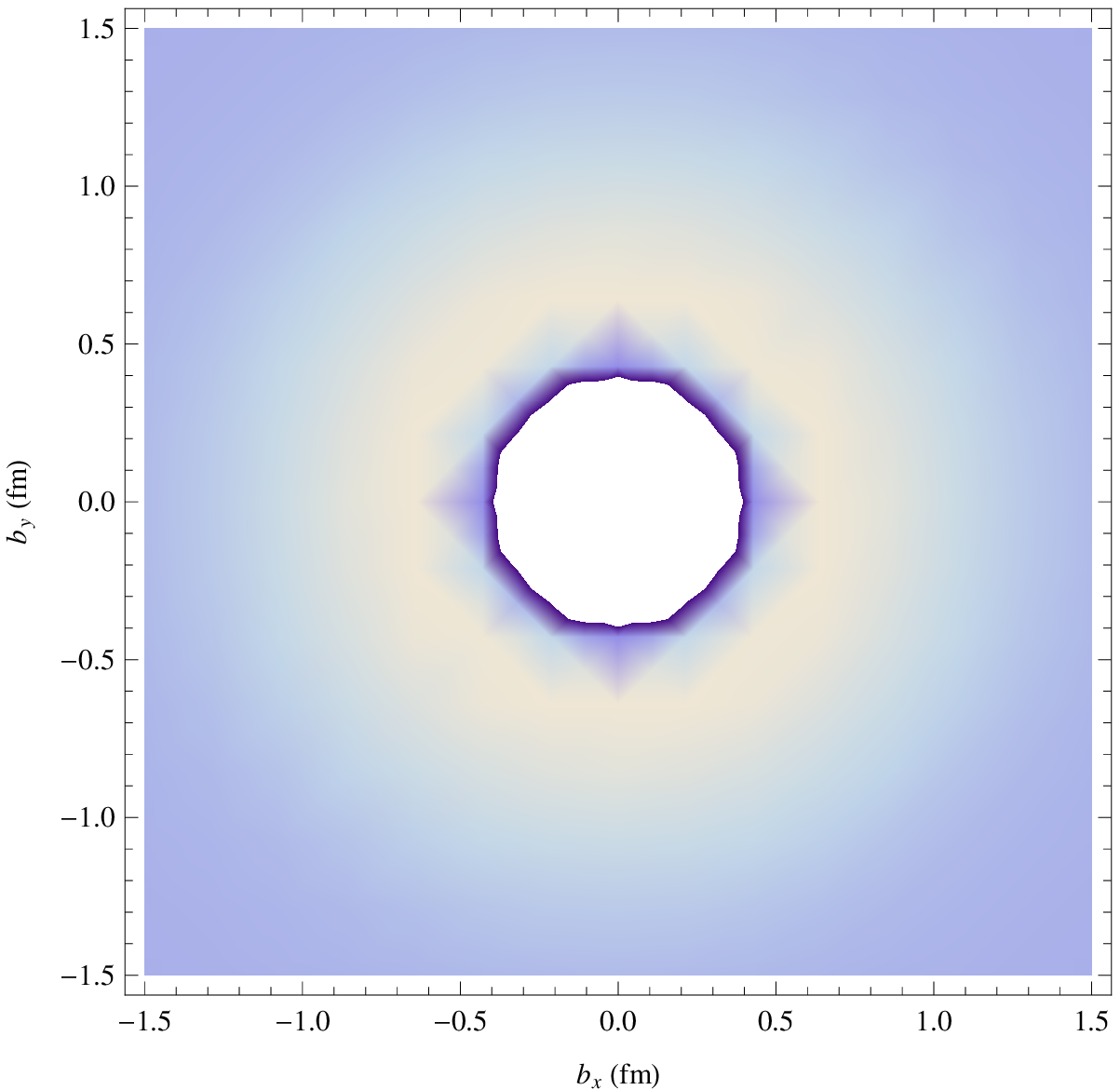,scale=0.6}
\end{center}
{\bf Fig.13:} Charge density $\rho_0^n(b_x,b_y)$. 

\vspace*{.5cm}

\begin{center}
\vspace*{1.25cm}
\epsfig{figure=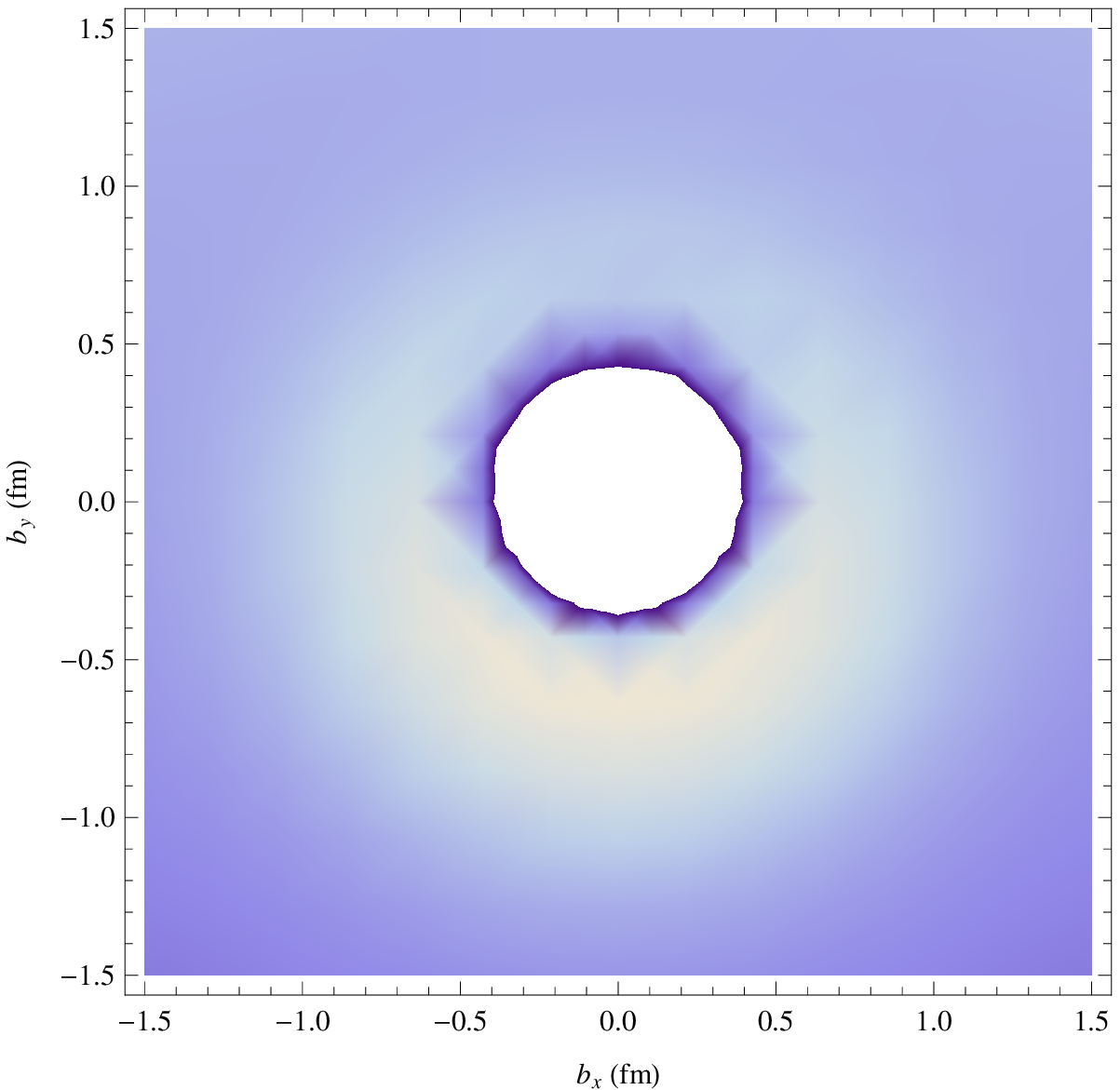,scale=0.6}
\end{center}
{\bf Fig.14:} Charge density $\rho_T^n(b_x,b_y)$. 
\end{figure} 

\newpage 

\begin{figure} 
\begin{center}
\vspace*{1.25cm}
\epsfig{figure=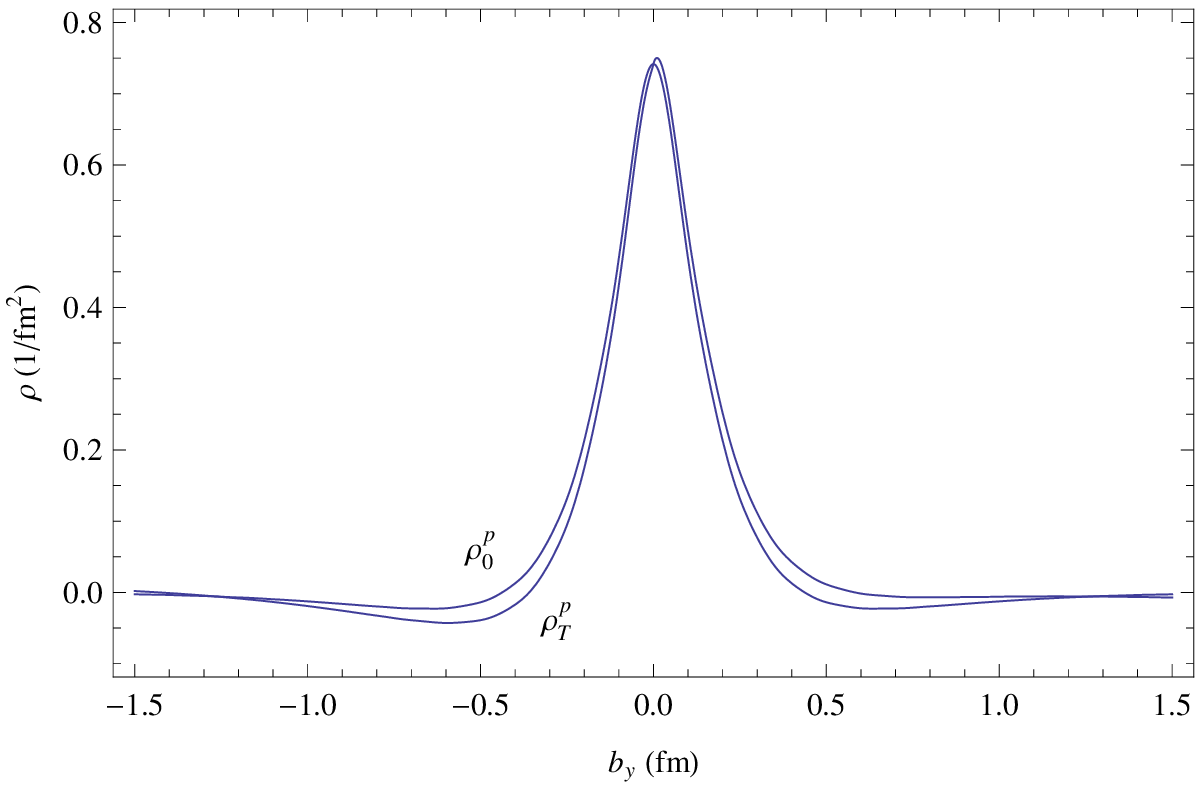,scale=0.6}
\end{center}
{\bf Fig.15:} Charge densities $\rho_0^p(b_y)$
and $\rho_T^p(b_y)$. 

\vspace*{.5cm}

\begin{center}
\vspace*{1.25cm}
\epsfig{figure=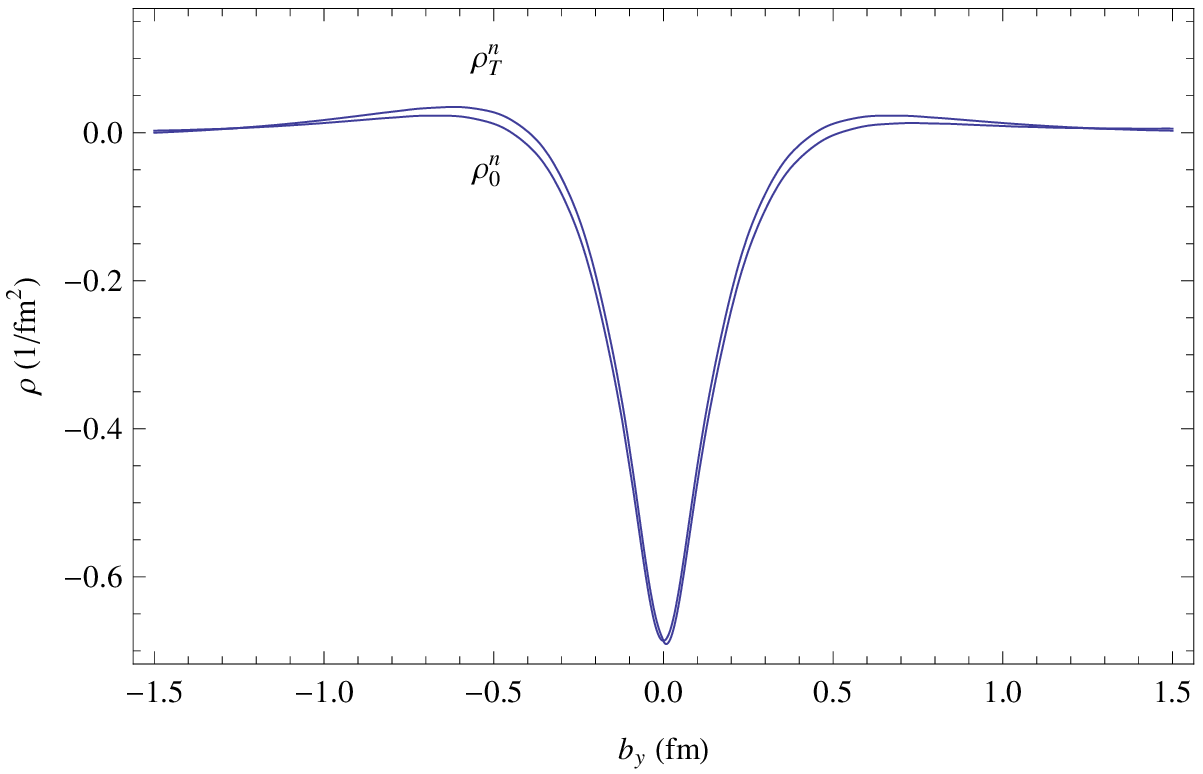,scale=0.6}
\end{center}
{\bf Fig.16:} Charge densities $\rho_0^n(b_y)$
and $\rho_T^n(b_y)$. 
\end{figure} 

\newpage 

\begin{figure} 
\begin{center}
\vspace*{1.25cm}
\epsfig{figure=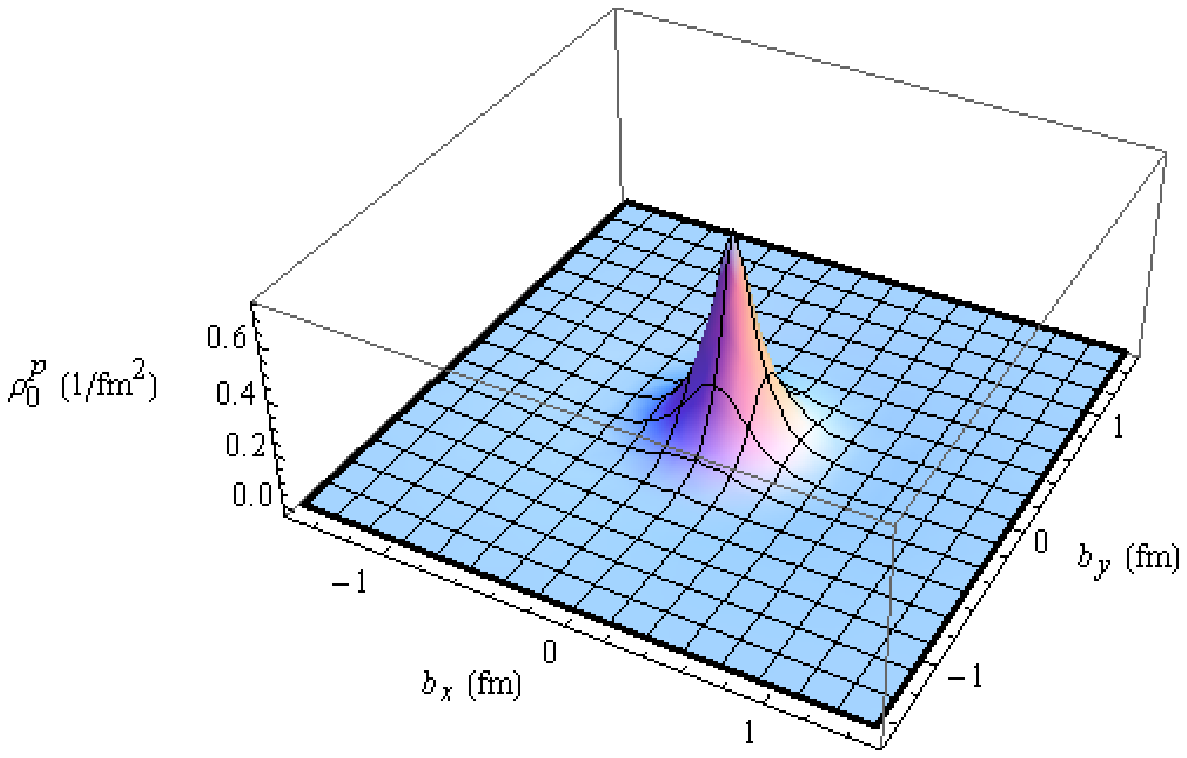,scale=0.6}
\end{center}
{\bf Fig.17:} 3D image of $\rho_0^p(b_x,b_y)$. 

\vspace*{.5cm}

\begin{center}
\vspace*{1.25cm}
\epsfig{figure=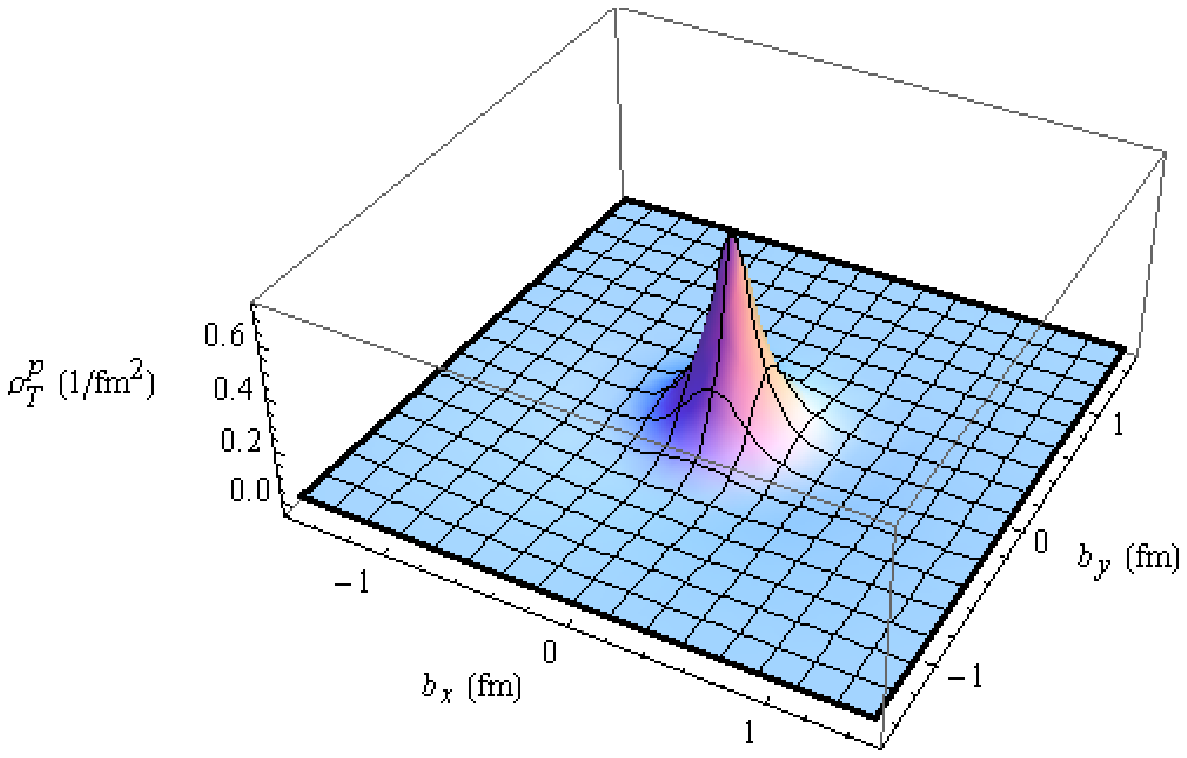,scale=0.6}
\end{center}
{\bf Fig.18:} 3D image of $\rho_T^p(b_x,b_y)$. 
\end{figure} 

\newpage 

\begin{figure} 
\begin{center}
\vspace*{1.25cm}
\epsfig{figure=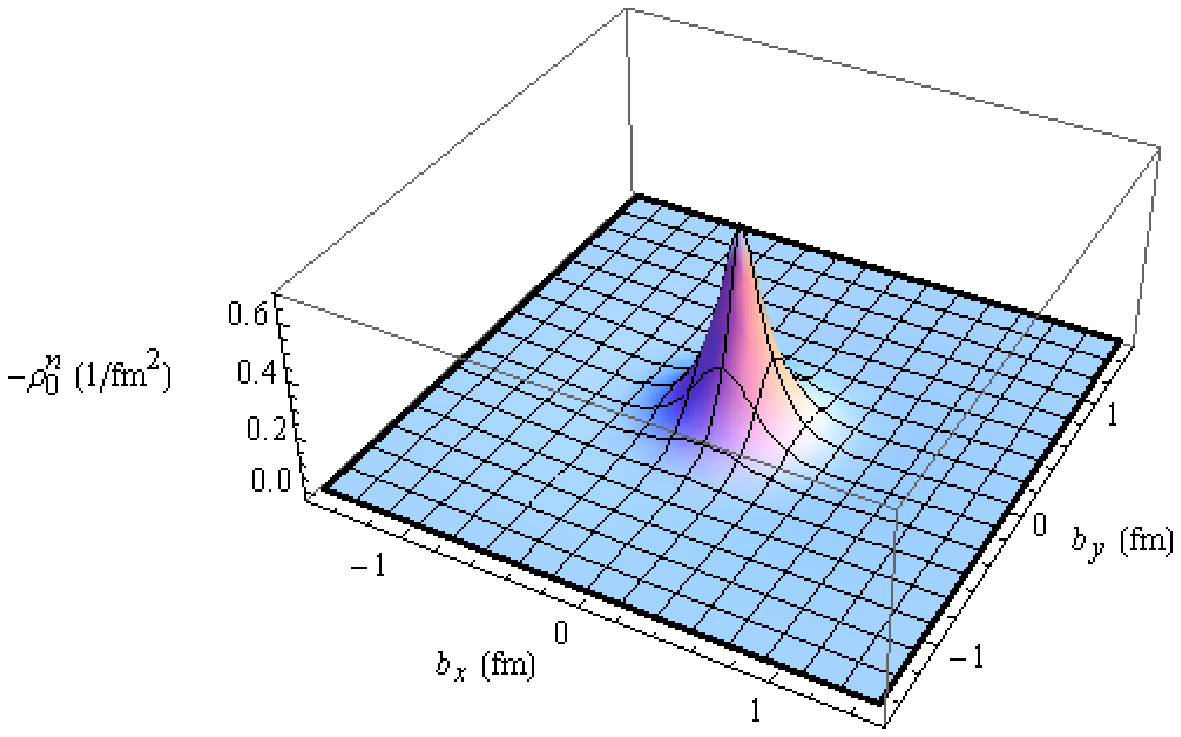,scale=0.6}
\end{center}
{\bf Fig.19:} 3D image of $-\rho_0^n(b_x,b_y)$. 

\vspace*{.5cm}

\begin{center}
\vspace*{1.25cm}
\epsfig{figure=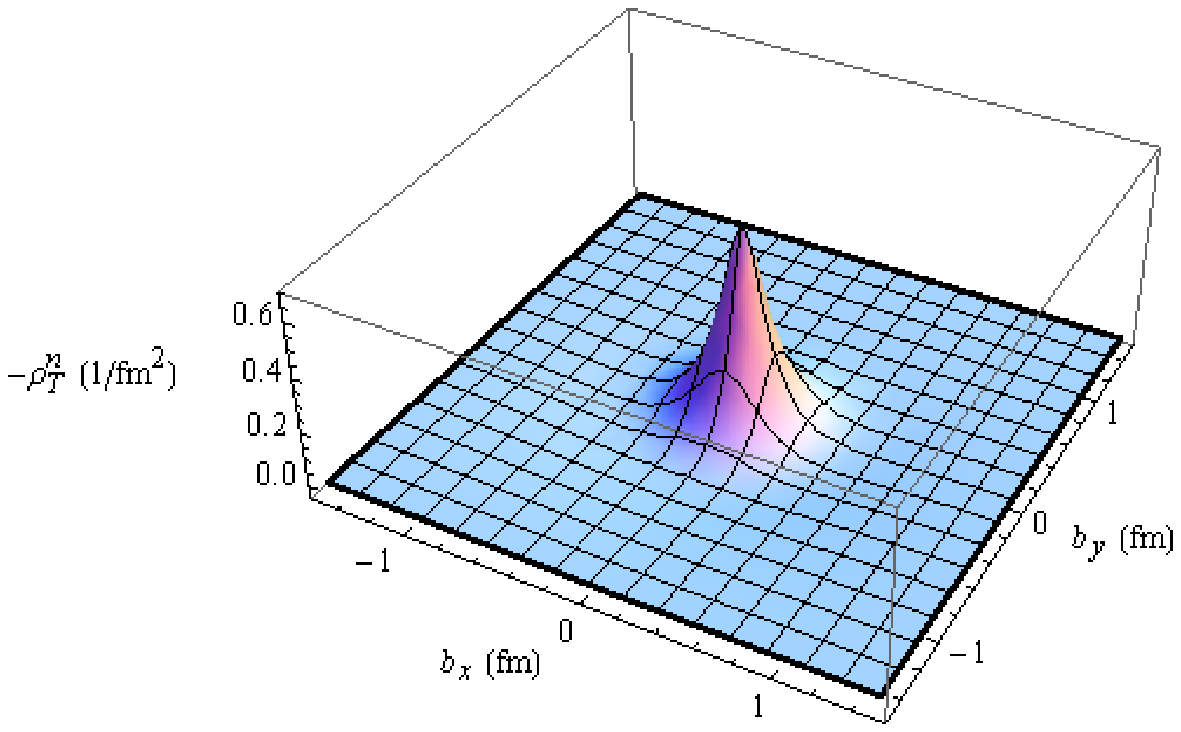,scale=0.6}
\end{center}
{\bf Fig.20:} 3D image of $-\rho_T^n(b_x,b_y)$. 
\end{figure} 

\end{document}